\documentclass{article}
\pdfoutput=1
\usepackage[utf8]{inputenc}
\usepackage{url}
\usepackage{amsmath}
\usepackage{amssymb}
\usepackage{mathtools}
\usepackage{graphicx}
\usepackage{grffile} 
\usepackage[shortlabels]{enumitem}
\usepackage{algorithm}
\usepackage{setspace}
\usepackage{algorithmic}
\usepackage{makecell, booktabs, array}

\usepackage{chngpage}

\usepackage{tikz}

\usetikzlibrary{positioning, calc, shapes.geometric, shapes, shapes.multipart, arrows.meta, arrows, decorations.markings, external, trees}\usepackage[round]{natbib}
\usepackage[margin=1in]{geometry}
\usepackage{amsmath}
\usepackage{amssymb}
\usepackage{amsthm}
\usepackage{afterpage}

\usepackage{mathtools}
\usepackage{chngpage, comment}

\newtheorem{theorem}{Theorem}

\title{\vspace{-1em}Adjusting for Partial Compliance in SMARTs: a Bayesian Semiparametric Approach}
\author{William J Artman\\[0.1cm]Department of Biostatistics and Computational Biology,\\ University of Rochester Medical Center,\\
Rochester, Saunders Research Building, 265 Crittenden Blvd., NY 14642, USA\\[0.1cm]
\url{William_Artman@URMC.Rochester.edu}\\[0.3cm] Ashkan Ertefaie\\[0.1cm]Department of Biostatistics and Computational Biology,\\ University of Rochester Medical Center,\\
Rochester, Saunders Research Building, 265 Crittenden Blvd., NY 14642, USA\\[0.3cm] Kevin G Lynch\\[0.1cm]
Center for Clinical Epidemiology and Biostatistics (CCEB)\\ and Department of Psychiatry,\\ University of Pennsylvania,\\ 3535 Market Street, 5099, Philadelphia, PA 19104, USA \\[0.3cm] James R McKay\\[0.1cm]Department of Psychiatry, Perelman School of Medicine,\\University of Pennsylvania,\\ 3535 Market St.,
Suite 500, Philadelphia, PA 19104, USA\\[0.3cm]
Brent A Johnson\\[0.1cm]Department of Biostatistics and Computational Biology,\\ University of Rochester Medical Center,\\
Rochester, Saunders Research Building, 265 Crittenden Blvd., NY 14642, USA}
\date{}
\begin{document}
\maketitle

\newcommand{\expit}{\mathrm{expit}}
\newcommand{\bD}{\boldsymbol D}
\newcommand{\bW}{\boldsymbol W}
\newcommand{\bZ}{\boldsymbol Z}
\newcommand{\bY}{\boldsymbol Y}
\newcommand{\bX}{\boldsymbol X}
\newcommand{\bx}{\boldsymbol x}

\newcommand{\bu}{\boldsymbol u}
\newcommand{\bgamma}{\boldsymbol \gamma}
\newcommand{\btheta}{\boldsymbol \theta}
\newcommand{\bbeta}{\boldsymbol \beta}
\newcommand{\bSigma}{\boldsymbol \Sigma}
\newcommand{\E}{\mathbb{E}}
\newcommand{\EDTR}{\mathrm{EDTR}}

\newcommand{\var}{\mathrm{Var}}
\newcommand{\indep}{\rotatebox[origin=c]{90}{$\models$}}
\singlespacing
\begin{abstract}

The cyclical and heterogeneous nature of many substance use disorders highlights the need to adapt
the type or the dose of treatment to accommodate the specific and changing needs of individuals. The Adaptive Treatment for Alcohol and Cocaine Dependence study (ENGAGE) is a multi-stage randomized trial that aimed to provide longitudinal data for constructing treatment strategies to improve patients' engagement in therapy. However, the high rate of noncompliance and lack of analytic tools to account for noncompliance have impeded researchers from using the data to achieve the main goal of the trial. We overcome this issue by defining our target parameter as the mean outcome under different treatment strategies for given potential compliance strata and propose a Bayesian semiparametric model to estimate this quantity. While it adds substantial complexities to the analysis, one important feature of our work is that we consider partial rather than binary compliance classes which is more relevant in longitudinal studies. We assess the performance of our method through comprehensive simulation studies. We illustrate its application on the ENGAGE study and demonstrate that the optimal treatment strategy depends on compliance strata.
\end{abstract}

\section{Introduction}

The cyclical and heterogeneous nature of substance use disorders highlights the need to tailor the type and/or the dose of treatment to the specific and changing needs of individuals (\citealp{mckay2009treating, kranzler2012personalized, black2014mechanisms, witkiewitz2015recommendations}). One important feature of substance use disorders research that makes constructing effective treatment regimes challenging is the high rate of noncompliance ($\geq 50\%$) to treatments. In general, failing to properly adjust for noncompliance can limit the usefulness and generalizability of the results.

A dynamic treatment regime is a treatment design that seeks to accommodate patient heterogeneity
in response to treatments \citep{murphy2001marginal,murphy2003optimal,robins2004optimal, chakraborty2013statistical,chakraborty2014dynamic,laber2014dynamic}. In dynamic treatment regimes, the type and/or dose of the treatment is adapted over time according to the
patient's characteristics and progress in treatment. The optimal dynamic treatment regime is one that optimizes the expected health
outcome of interest. 
Recently, there has been increased interest in sequential, multiple assignment, randomized trials (SMARTs),
which were specifically developed to provide empirical evidence for the construction of optimal dynamic treatment regimes. The
SMART is a clinical trial design in which the randomization scheme takes into account patient response to prior
treatments \citep{LAVORI2000605,murphy2005experimental, lei2012smart,nahum2012experimental,chakraborty2013statistical}. Several sequences of treatments of scientific interest are embedded in a SMART to form the
embedded dynamic treatment regimes. These often use a single tailoring variable, such as the individuals' early response
status.

Existing methods for analyzing SMART data are limited to intention-to-treat analyses. Intention-to-treat analysis measures the effect of randomization to a particular group on the outcome regardless of whether the individual complied with their assigned treatment. There are two main criticisms of intention-to-treat analyses: 1) the
treatment effect estimates are biased due to confounding; the bias is often toward the null hypothesis of no
effect because the estimate occurs across individuals with different levels of compliance \citep{marasinghe2007noncompliance, lin2008longitudinal}. In fact, the intention-to-treat
effect estimate will likely diminish as noncompliance increases; 2) the rate of compliance or the compliance
pattern in standard clinical practice may not be the same as the rate or the pattern in the clinical trial \citep{frangakis1999addressing, robins1991correcting, hewitt2006there}.
These two shortcomings can seriously limit the usefulness and generalizability of the concluding results. Noncompliance
rates are particularly high in substance use disorder treatment (\citealp{us1992substance}). This is a major challenge for the Adaptive Treatment for Alcohol and Cocaine Dependence (ENGAGE) study, which attempts to deliver new interventions to patients who have failed in the first treatment
administered and therefore are at risk to either not initiate the new intervention or to drop out quickly \citep{mckay2015effect,van2015treatment}.

\subsection{Accounting for noncompliance}

An important challenge in adjusting for compliance is that observed compliances are post-treatment variables. Therefore, conditioning on them may induce the post-treatment adjustment bias. To overcome this issue, instrumental variable and principal stratification based methods have been proposed. Instrumental variable analyses are widely used to adjust for binary compliances in two-arm randomized trials. In this case, treatment assignment may be used as an instrument to obtain an unbiased estimate of the treatment effect for compliers where compliers are the subgroup of patients who would receive the treatment that matches the instrumental variable regardless of the instrumental variable's value  (\citealp{angrist1995two}; \citealp{angrist1996identification};
\citealp{angrist2006instrumental}; \citealp{wooldridge2010econometric}; \citealp{wooldridge2015introductory}; \citealp{ertefaie2018discovering}).
 
In the context of individualized medicine, \cite{cui2019semiparametric} developed a semiparametric approach to account for noncompliance. In particular, they provide a method for non-parametric identification of the optimal treatment strategy among compliers for a single-stage trial in the presence of binary noncompliance. However, in longitudinal studies, compliance is often defined as the average compliance  measured through the follow-up time which is a continuous variable. In this case, in order to use instrumental variable based approaches with a binary instrument, the partial compliance must be thresholded to form a binary compliance variable. This procedure has two shortcomings: (1) the results may depend on the threshold used for dichotomization; and (2) it leads to loss of information in compliance behaviour due to the dichotomization.

Principal stratification provides an alternative framework to account for noncompliance (\citealp{efron1991compliance,frangakis2002principal,frangakis2002clustered,greevy2004randomization}).  \cite{jin2008principal} employed a Bayesian approach to principal stratification by modeling partial potential compliances using the beta distribution. Their method relies on a monotonicity assumption which states placebo compliance is always higher than treatment compliance. This assumption may not be reasonable for randomized clinical trials with two active arms where there is not a clear ordering of potential compliances. Moreover, their method does not allow modeling complicated distributions of compliance for which the beta distribution is a poor fit. \cite{schwartz2011bayesian} proposed an alternative Bayesian semiparametric framework to relax some of the restrictive assumptions made in \cite{jin2008principal}. Specifically, \cite{schwartz2011bayesian} estimated the joint distribution of the potential compliances non-parametrically using a Dirichlet process (DP) mixture and data augmentation to impute the latent potential compliances \citep{kim2019longibayesian, kim2019bayesian}. The use of a DP mixture for kernel density estimation allows complicated distributions of compliance to be modeled giving it an advantage over other methods by relaxing the parametric assumptions made in \cite{jin2008principal}. It furthermore clusters individuals which allows information to be shared between observations within clusters aiding in estimation of the joint potential compliance distribution. 
Frequentist approaches have also been explored to account for partial compliance in randomized clinical trials.  \cite{bartolucci2011modeling} proposed a principal stratification based method where the joint distribution of potential compliances were modeled using a Plackett copula. This approach has the disadvantage of empirical findings depending on the choice of a sensitivity parameter specifying association in the Plackett copula. 
The existing methods, focus entirely on single time point treatment assignment settings and are not applicable to SMARTs. Our method fills this gap.

\subsection{Our contribution}

We propose a method to estimate the mean outcome under a given treatment strategy in the presence of partial compliance. We leverage principal stratification in which strata are formed by \textit{potential} compliances. The potential compliance of a particular treatment is the compliance that would have been observed, had the patient been assigned to that treatment. These are pre-treatment patient characteristics; hence, they may be treated as baseline covariates. We refer to the principal strata as compliances classes. An important challenge with principal stratification is that some of the potential compliances are latent. 

We propose a Bayesian semiparametric approach for estimating the mean treatment strategy outcome given compliance classes. A crucial step for identifiability of our target parameter is to estimate the joint distribution of the potential compliances which is used to augment the unobserved potential compliances. To reduce the chance of model misspecification, we estimate the joint distribution non-parametrically using a Dirichlet process mixture.  
 The method accounts for treatment effect heterogeneity with respect to patient's compliance behavior, which in turn leads to finding optimal dynamic treatment regimes that better suit the patients' needs/desires and boosts the interpretability of the results. The proposed method builds upon \cite{schwartz2011bayesian} and extends the existing principal stratification methods from single-stage to multi-stage treatment decision settings. Furthermore, for a given compliance class, we determine which dynamic treatment regimes are indistinguishable from the optimal dynamic treatment regime within a class of regimes using multiple comparisons with the best (MCB) \citep{hsu1981simultaneous,hsu1996multiple}. The best refers to the embedded DTR which maximizes some health outcome. MCB adjusts for multiple comparisons and yields greater power compared with methods which entail all pairwise comparisons (\citealp{ertefaie2015identifying}; \citealp{artman2018power}). The interpretation of the set of best dynamic treatment regimes is appealing to investigators as it sheds light on how the optimal dynamic treatment regimes change depending on compliance class while taking into account the uncertainty of each dynamic treatment regime outcome.

\subsection{Outline }
 In Section 2, we summarize the ENGAGE SMART. In Section 3, we introduce our framework including notation, the estimand, and the likelihood. Next, in Section 4, we present the non-parametric Bayesian model for partial compliance in full generality. Afterward, we describe the estimation and inference procedure using a Gibbs sampler in Section 5. In Section 6, we present our assumptions in full generality as well as for the two SMART examples. In this Section, we present a theorem which proves identifiability of the regression coefficients in the outcome-potential compliance model subject to our assumptions. After we finish discussing estimation of the embedded DTRs, we transition to Section 7, where we discuss model selection. We subsequently summarize multiple comparisons with the best in Section 8 which is a procedure for comparing embedded DTRs. After we fully describe our methodology, we examine the performance of our method using extensive simulation studies in Section 9. In Section 10, we then apply our method to the real ENGAGE SMART study. Lastly, we conclude with a discussion in Section 11. Omitted proofs and figures can be found in the Appendix.

\section{ENGAGE SMART study} \label{sec:engage}

\begin{figure}[t]
\centering
\includegraphics[width = 5.5in, trim = {6cm 5cm 5cm 2cm},clip=true]{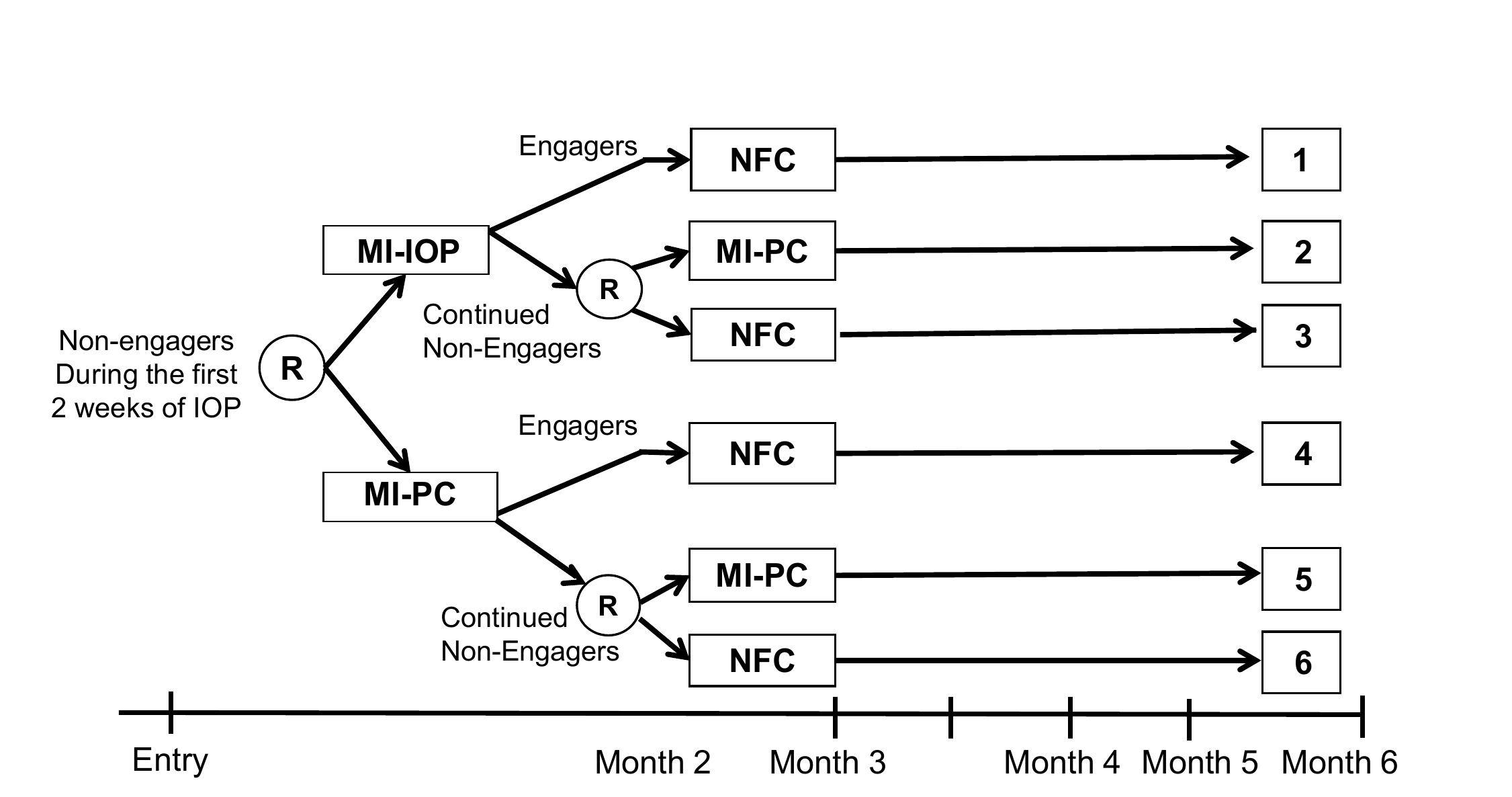}
\caption{Diagram of the ENGAGE SMART.}
\label{fig:ENGAGE-Diagram}
\end{figure}
 \begin{figure}[t]
\centering
\includegraphics[width = 7in, trim = {0cm 2cm 0cm 2cm},clip=true]{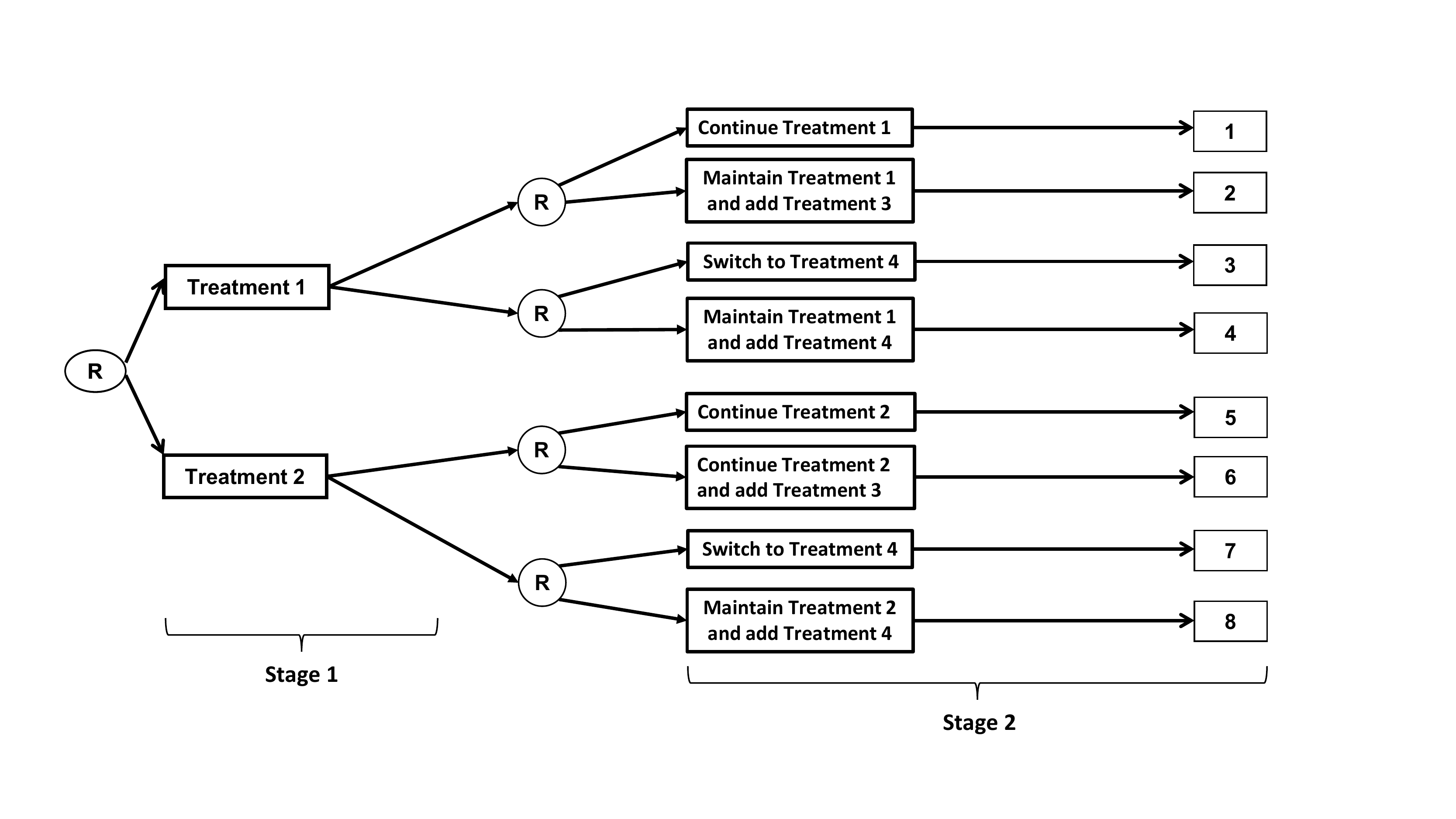}
\caption{Diagram of the General SMART.}
\label{fig:General-SMART}
\end{figure}

The subjects in the Adaptive Treatment for Alcohol and Cocaine Dependence (ENGAGE) study were patients who abuse alcohol and patients who abuse cocaine (\citealp{mckay2015effect,van2015treatment}). The primary goal of the ENGAGE study was to determine the effect of permitting patients who failed to engage in or dropped out of intensive outpatient programs (IOPs) to choose their subsequent treatment.
The study initially followed patients who were in IOP. After 2 weeks, those who failed to engage in IOP were randomized to one of two motivational interviewing (MI) interventions. Non-engagement was defined as  failing to attend two or more IOP session in Week 2. The first intervention encouraged patients to engage in IOP (MI-IOP). The second offered patients a choice for treatment (called patient choice or MI-PC). For our analysis, we consider the study from the 2 week point where patients were first randomized. PC consisted of choosing from cognitive behavioral therapy, telephone-based stepped care, IOP, or medical management. After 8 weeks, subjects who failed to engage in IOP (defined as not attending any IOP sessions in Weeks 7 and 8) were re-randomized to either MI+PC or no further outreach care (NFC).  See Figure \ref{fig:ENGAGE-Diagram} for a diagram depicting the SMART. Note that not-engaging is related to, but not the same as not complying. Compliance is the fraction of sessions attended while engaging is meeting certain criterion defined in the study which is used as a tailoring variable.

 In this study, there were three potential compliances corresponding to MI-IOP stage 1, MI-PC stage 1, and MI-PC stage 2. There were 4 embedded DTRs. Two of the embedded DTRs were as follows:
\begin{enumerate} 
\item Start with MI+IOP. If the subject is engaged 
during the first 8 weeks, then at the 8-week point, offer NFC; if the subject is labeled as a non-engager at week 8 of follow-up, offer MI+PC. 
\item Start with MI+IOP. If the subject is engaged during the first 8 weeks, then at the 8-week point, offer NFC; if the subject is labeled as a non-engager at week 8 of follow-up, offer NFC. 
\end{enumerate}
The other two embedded DTRs are similar (see Table \ref{tab:EDTR-ENGAGE-table}). See \cite{mckay2015effect,van2015treatment} for more details about the SMART.

\begin{figure}[t]
\centering
\includegraphics[width = 2in,trim = {0cm 0cm 0cm 0cm},clip=true]{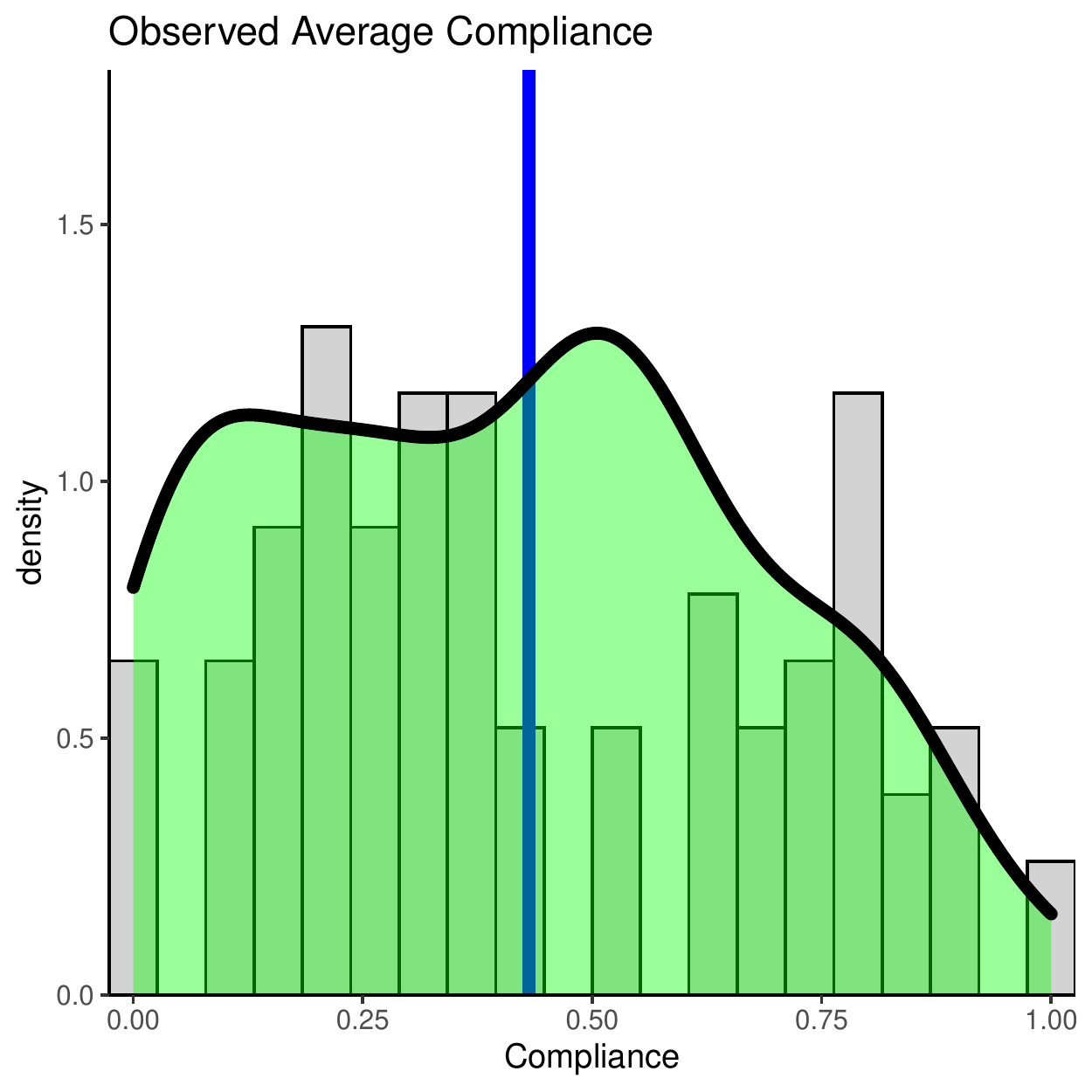}
\caption{Overall compliance histogram. Vertical line is median observed compliance.}
\label{fig:compliance-distribution}
\end{figure}

\begin{figure}[t]
\centering
\includegraphics[width = 3in, trim = {0cm 0cm 0cm 0cm},clip=true]{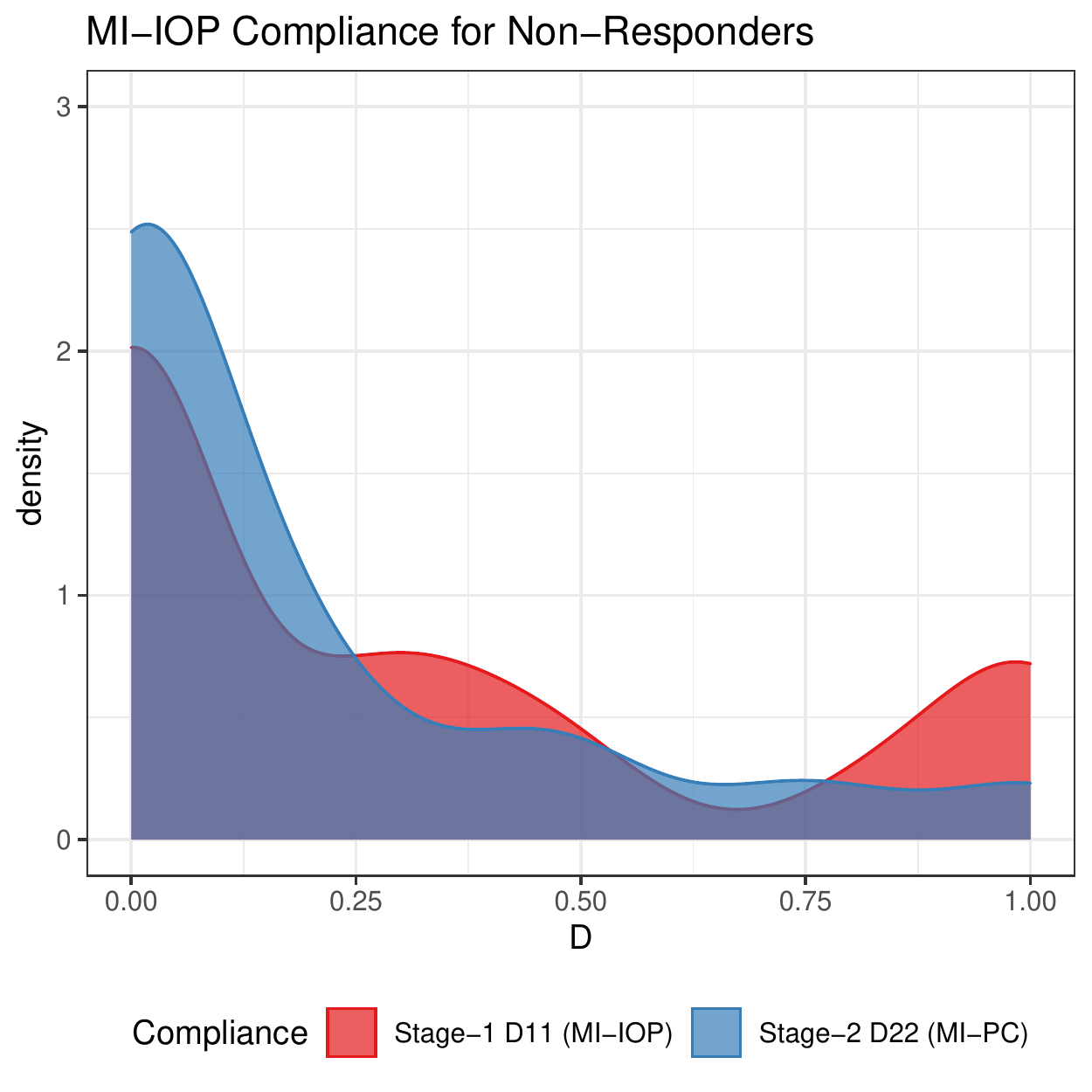}
\includegraphics[width = 3in, trim = {0cm 0cm 0cm 0cm},clip=true]{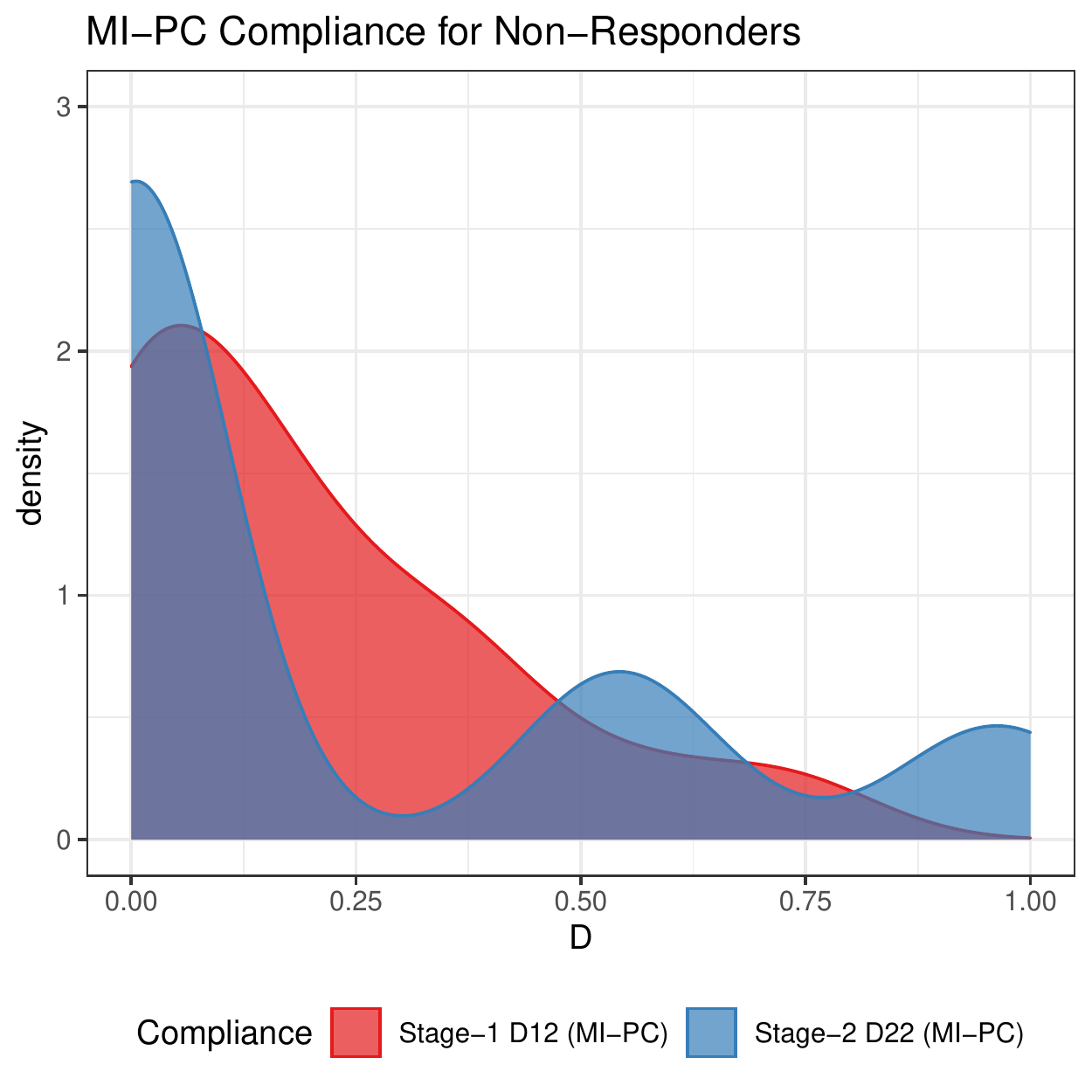}
\caption{Non-responder compliance density plots for each stage.}
\label{fig:Compliance-densities}
\end{figure}

Figure \ref{fig:compliance-distribution} shows the low overall compliance in the ENGAGE study emphasizing the importance of incorporating compliance information when comparing embedded DTRs. Figure \ref{fig:Compliance-densities} shows the distribution of the observed compliance for non-responders to stage-1 treatment MI-IOP on the left and MI-PC on the right. Stage-2 compliance to MI-PC is overlayed. Note that the compliance for stage-2 is substantially lower than that of stage-1 treatments. This is what one might expect since stage-2 subjects are non-engagers and hence low compliers. The stage-1 treatment observed compliances for MI-IOP and MI-PC are similar; however, there is more weight near 100\% for MI-IOP compared with MI-PC stage-1. 

Determination of optimal embedded DTR in ENGAGE would inform physicians about whether permitting patients to choose their care is superior to allowing physicians to choose subsequent care. The outcome included measures of alcohol and cocaine use between week 2 and 24. We define partial compliance as the fraction of sessions attended for an intervention. 
One might expect that compliance is lower when physicians choose leading to poorer outcomes. By allowing patients choice of care, outcomes may be improved as compliance will be higher. However, if a patient is known a priori to be a high complier, the physician's assignment may be superior to patient's choice due to greater subject matter knowledge of the physician compared with the patient. 

\begin{figure}[t]
\centering
\includegraphics[width = 6in, trim = {0.0cm 0.0cm 0.0cm 0.0cm},clip=true]{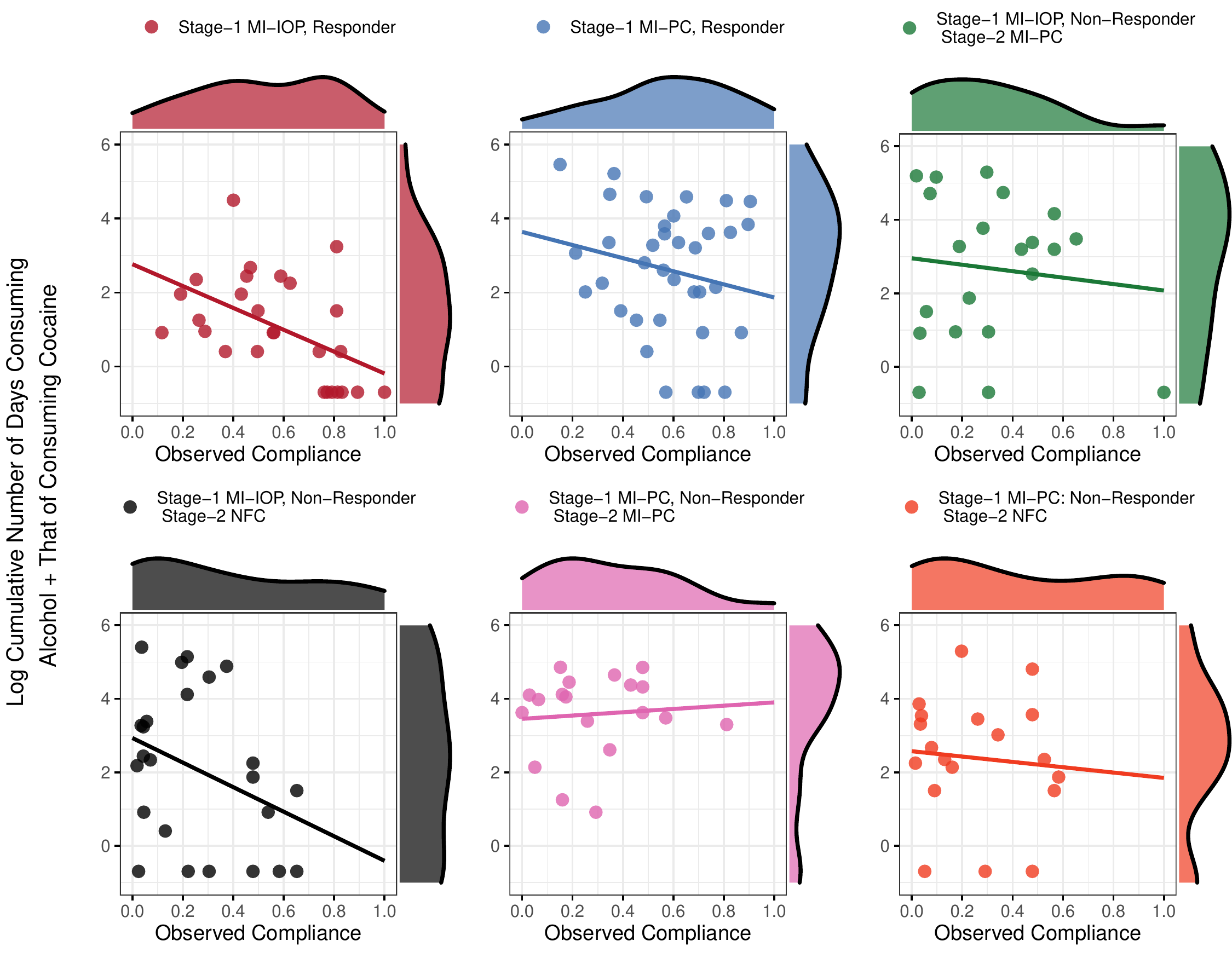}
\caption{Log of the cumulative number of days of consuming alcohol and cocaine vs. observed compliance with marginal densities.}
\label{fig:outcome-vs.-compliance}
\end{figure}

In Figure \ref{fig:outcome-vs.-compliance}, we plot the log of the cumulative number of days of drinking alcohol and using cocaine vs. observed compliances for each of the six treatment sequences in ENGAGE shown in Figure \ref{fig:ENGAGE-Diagram}. Higher values imply poorer outcomes. We see a negative trend for all groups except for the sequence Stage-1 MI-PC, non-responder, stage-2 MI-PC.  The top marginal density plot shows that observed compliance is higher for MI-PC which makes sense as patients get to choose their intervention.  The interpretation of Figure \ref{fig:outcome-vs.-compliance} should be taken with care as the compliances are observed rather than potential.

\section{Framework}
\subsection{Notation }

We develop our method under a more general SMART structure than the ENGAGE study by allowing both responders and non-responders to stage-1 interventions to be re-randomized at the second stage (Figure \ref{fig:General-SMART}).
Let $A_1\in\{-1,+1\}$ be the stage-1 treatment indicator, $A_2^{\mathrm{R}}\in\{-1,+1\}$ be the stage-2 treatment indicator for responders to the stage-1 treatment, and $A_2^{\mathrm{NR}}\in\{-1,+1\}$ be the stage-2 treatment indicator for non-responders to the stage-1 treatment. 
 Let $D_{11i}$ denote the stage-1 potential compliance for subject $i$ had they been randomized to $A_{1i}=+1$ 
and $D_{12i}$ denote the stage-1 potential compliance for subject $i$ had they been randomized to $A_{1i}=-1$. 
Similarly, define $D_{2j}^{\mathrm{R}}$ and $D_{2j}^{\mathrm{NR}}$ as the $j^{\mathrm{th}}$ ($j=1,2$) stage-2 potential compliances for responders and non-responders to stage-1 treatment, respectively. We assume that the partial compliances take their value in $\mathcal{R}^{[0,1]}$ (i.e., $(D_{11},D_{12},D_{21}^{\mathrm{R}},D_{22}^{\mathrm{R}},D_{21}^{\mathrm{NR}},D_{22}^{\mathrm{NR}}) \in [0,1]^6$). Let $S_i$ denote the observed stage-1 treatment response indicator for subject $i$. Let $Y_{ik}$, $k=1,...,K$ denote the potential outcome of the $k^{\mathrm{th}}$ treatment sequence at the end of the study for subject $i$. Let $Y_i^{(l)}$, $l=1,...,L$ denote the potential outcome of the $l^{\mathrm{th}}$ embedded DTR for subject $i$. In ENGAGE, $K=6$ and $L=4$ while in the general SMART (Figure \ref{fig:General-SMART}), $K=L=8$. Let $n$ be the total sample size in the SMART.

 The general SMART includes 8 embedded dynamic treatment regimes listed in Table \ref{tab:generalsmartEDTR}. There are a total of $6$ possible potential compliances: two for stage 1 ($D_{11},D_{12}$); two for responders to stage-1 treatments ($D_{21}^{\mathrm{R}},D_{22}^{\mathrm{R}}$); and two for non-responders to stage-1 treatments ($D_{21}^{\mathrm{NR}},D_{22}^{\mathrm{NR}}$). Because, the first stage-2 intervention option for responders is the same intervention as for stage 1, it is plausible to assume that $D_{21}^{\mathrm{R}}=D_{11}$. This reduces the number of potential compliances to $5$ (i.e., $D_{11},D_{12},D_{22}^{\mathrm{R}},D_{21}^{\mathrm{NR}},D_{22}^{\mathrm{NR}}$). In the ENGAGE study, because there is no re-randomization among responders, the potential compliances $D_{21}^{\mathrm{R}}$ and $D_{22}^{\mathrm{R}}$ do not exist. Moreover, in ENGAGE, non-responders were re-randomized to a placebo (i.e., NFC) and an active intervention which reduces the number of potential compliances among non-responders. Specifically, because patients assigned to NFC do not have access to the other treatment option the only potential compliance for non-responders is $D_{22}^{\mathrm{NR}}$. Thus, in the ENGAGE study, there are only three potential compliances: $D_{11},D_{12}$, and $D_{22}(\equiv D_{22}^{\mathrm{NR}}$).

\subsection{Estimand}

The estimand of interest is the average outcome for a particular embedded DTR $Y_i^{(l)}$, in a compliance class specified by $\boldsymbol{D}$. These are known as the \textit{principal causal effects} and are defined as $\mathrm{PCE}^{(l)}(\bD)=\E[Y^{(l)}\mid \boldsymbol{D}],$ for $l=1,2,\cdots,L$ where $L$ is the number of embedded DTRs.

In the SMART setting, the principal causal effects can also be represented as
\begin{align*}
\mathrm{PCE}^{(l)}(\bD)=\E[Y\mid A_{1}=a_{1l}, &S=1,A_2^{\mathrm{R}}=a_{2l},\boldsymbol{D}]\Pr(S=1\mid A_1=a_{1l},\boldsymbol{D})+\\
&\E[Y\mid A_1=a_{1l},S=0,A_2^{\mathrm{NR}}=a_{2l},\boldsymbol{D}]\Pr(S=0\mid A_1=a_{1l},\boldsymbol{D}), \hspace{.1in} l=1,2,\cdots,L,
\end{align*}
where $Y$ is the observed outcome and $a_{jl}$ is the treatment option at stage $j$ ($j=1,2$) that is consistent with the $l^{\mathrm{th}}$ embedded DTR (see Appendix A). Although, for illustrative purposes, we only focus on unconditional principal causal effects (i.e., averaged over the baseline covariates), the proposed methods can be readily applied to principal causal effects conditional on baseline covariates as well. The main challenge in estimating the principal causal effects is that the compliance classes are partially latent. We propose a method to impute the missing potential compliances leveraging a Bayesian non-parametric model and a parametric outcome model. The procedure is presented in the next section. We now describe how to leverage the PCE representation with Markov Chain Monte Carlo.

Let $\theta(\boldsymbol{D})^{(l)}=\E[Y^{(l)}\mid \boldsymbol{D}]$, $\theta_{\mathrm{R}}(\boldsymbol{D})=\E[Y\mid A_{1}=a_{1l}, S=1,A_2^{\mathrm{R}}=a_{2l},\boldsymbol{D}]$, $\theta_{\mathrm{NR}}(\boldsymbol{D})=\E[Y\mid A_1=a_{1l},S=0,A_2^{\mathrm{NR}}=a_{2l},\boldsymbol{D}]$, and $\lambda_{A_1}(\boldsymbol{D})=\Pr(S=1\mid A_1=a_{1l},\boldsymbol{D})$.
Given a vector of augmented potential compliances $\boldsymbol{D}$, at the $m^{th}$ iteration of the Gibbs sampler, we draw $\theta_{\mathrm{R}}^{(m)}(\boldsymbol{D})$, $\theta^{(m)}_{\mathrm{NR}}(\boldsymbol{D})$, and  $\lambda_{A_1}^{(m)}(\boldsymbol{D})$.
Hence, the $m^{th}$ draw of the principal causal effect is $$\theta(\boldsymbol{D})^{(l),(m)}=\theta_{\mathrm{R}}(\boldsymbol{D})^{(m)}\lambda_{A_1}^{(m)}(\boldsymbol{D})+\theta_{\mathrm{NR}}^{(m)}(\boldsymbol{D})(1-\lambda_{A_1}^{(m)}(\boldsymbol{D}))$$

Then, by the law of large numbers, $\E[Y^{(l)}\mid \bD]\approx \dfrac{1}{M}\sum_{m=1}^{M}\theta(\boldsymbol{D})^{(l),(m)}$. 
Throughout, we fit the conditional means, $\theta_{\mathrm{R}}(\boldsymbol{D})^{(m)}$, using Bayesian linear regression and the probability of response, $\lambda_{A_1}^{(m)}(\boldsymbol{D})$, using Bayesian logistic regression.

We now describe the complete-data likelihood. Assume i.i.d. observations. Define $\btheta$ as the vector of parameters. We have that the complete-data likelihood is as follows:

\begin{align*}
&\Pr(\bY^{\mathrm{obs}},\bD,\boldsymbol{S}^{\mathrm{obs}},\boldsymbol{A}_1,\boldsymbol{A}_2\mid \btheta)\\
&\propto \prod_{i=1}^n \Pr(Y_i^{\mathrm{obs}},\bD_i, {S}_i^{\mathrm{obs}},A_{1i},A_{2i}\mid \btheta)\\
&\propto \prod_{i=1}^n \Pr(Y_i^{\mathrm{obs}}\mid \bD_i,S_i^{\mathrm{obs}},A_{1i},A_{2i},\btheta)\Pr(S_i^{\mathrm{obs}}\mid \bD_i,A_{1i}, \btheta)\Pr(\bD_i\mid\btheta)
\end{align*}

\section{Non-parametric Bayesian model for potential compliances}

 In this section we specify the Dirichlet process (DP) mixture model to estimate non-parametrically the joint distribution of the potential compliances. 
Consider the DP with base measure $G_0$ and concentration parameter $\alpha$ which specifies the sparsity of a discrete realization of the DP. The realization of a DP is itself a distribution. A random measure $G\sim \mathrm{DP}(\alpha G_0)$ may be constructed using the stick-breaking representation (\citealp{schwartz2011bayesian}): \begin{align}
    G(\cdot)=\sum_{h=1}^{\infty} w_h \delta_{\boldsymbol{\gamma}_h}(\cdot)&\text{ and }  \boldsymbol{\gamma}_h \overset{iid}{\sim} G_0,\nonumber\\ w_h=w_h'\prod_{k<h}(1-w'_k)&\text{ and }w_h' \overset{iid}{\sim} \mathrm{Beta}(1,\alpha).\nonumber
    \end{align} 
where $\delta_{\boldsymbol{\gamma}}(\cdot)$ is a point mass at $\boldsymbol{\gamma}$.
The joint distribution of the potential compliances is estimated as follows: \begin{equation}f_{\boldsymbol{D}}(\boldsymbol{d}\mid\bgamma)=\int K(\boldsymbol{d} \mid \boldsymbol{\gamma})dG(\boldsymbol{\gamma})\label{integralkde}\end{equation} where $G \sim \mathrm{DP}(\alpha G_0)$ and $K(\cdot\mid \boldsymbol{\gamma})$ is a kernel. Compliances are between $0$ and $1$ so we use the multivariate truncated normal distribution restricted to the unit (hyper)cube $[0,1]^m$ as a kernel where $m$ is the number of potential compliances. Then, (\ref{integralkde}) is equivalent to
\begin{table}[t]
\centering
\begin{tabular}[t]{cc}
\toprule
 & Embedded Dynamic Treatment Regime\\
\toprule
1 & $A_1=+1,S=1$ or $A_1=+1,S=0,A_2=+1$\\
&Receive MI-IOP, if respond receive NFC, if not respond switch to MI-PC\\
\midrule
2 & $A_1 = +1, S=1$ or $A_1=+1,S=0,A_2=-1$\\
&Receive MI-IOP, if respond, receive NFC, if not respond, receive NFC\\
\midrule
3 & $A_1=-1,S=1$ or $A_1=-1,S=0,A_2=+1$\\
&Receive MI-PC, if respond receive NFC, if not respond switch to MI-PC\\
\midrule
4 & $A_1=-1,S=1$ or $A_1=-1,S=0,A_2=-1$\\
&Receive MI-PC, if respond receive NFC, if not respond, receive NFC.\\
\bottomrule
\end{tabular}
\caption{Embedded dynamic treatment regime decision rules for the ENGAGE SMART.}
\label{tab:EDTR-ENGAGE-table}
\end{table}

 \begin{table}[t]
\centering
\begin{tabular}[t]{cc}
\toprule
 & Embedded Dynamic Treatment Regime\\
\toprule
1 & $A_1=+1,S=1,A_2^{\mathrm{R}}=+1$ or $A_1=+1,S=0,A_2^{\mathrm{NR}}=+1$\\
&Receive Trt. 1, if respond, continue trt 1, if not respond, switch to trt 4.\\
\midrule
2 & $A_1 = +1, S=1,A_2^{\mathrm{R}}=+1$ or $A_1=+1,S=0,A_2^{\mathrm{NR}}=-1$\\
&Receive Trt. 1, if respond continue Trt 1, if not respond, add trt. 4\\
\midrule
3 & $A_1=+1,S=1,A^{\mathrm{R}}_2=-1$ or $A_1=+1,S=0,A_2^{\mathrm{NR}}=+1$\\
&Receive trt 1, if respond add trt 3, if not respond switch to trt 4\\
\midrule
4 & $A_1=+1,S=1,A^{\mathrm{R}}_2=-1$ or $A_1=+1,S=0,A_2^{\mathrm{NR}}=-1$\\
& Receive Trt. 1, if respond add Trt 3., if not respond add trt. 4.\\
\midrule
5 & $A_1=-1,S=1,A_2^{\mathrm{R}}=+1$ or $A_1=-1,S=0,A_2^{\mathrm{NR}}=+1$\\
&Receive trt.2, if respond continue trt. 2., if no respond switch to trt 4.\\
\midrule
6 & $A_1 = -1, S=1,A_2^{\mathrm{R}}=+1$ or $A_1=-1,S=0,A_2^{\mathrm{NR}}=-1$\\
&Receive trt. 2, if respond, continue trt. 2, if not respond, add trt 4.\\
\midrule
7 & $A_1=-1,S=1,A^{\mathrm{R}}_2=-1$ or $A_1=-1,S=0,A_2^{\mathrm{NR}}=+1$\\
& Receive trt.2, if respond add trt. 3, if not respond switch to trt 4.\\
\midrule
8 & $A_1=-1,S=1,A^{\mathrm{R}}_2=-1$ or $A_1=-1,S=0,A_2^{\mathrm{NR}}=-1$\\
&Receive trt. 2, if respond add trt. 3., if not respond, add trt 4.\\
\bottomrule
\end{tabular}
\caption{Embedded dynamic treatment regime decision rules for the General SMART.}
\label{tab:generalsmartEDTR}
\end{table}

\begin{equation}f_{\boldsymbol{D}}(\boldsymbol{d}\mid\{\bgamma_h\})=\sum_{h=1}^{\infty}w_h c_h \mathcal{N}\left(\boldsymbol{d}\mid \boldsymbol\eta_{h},\bSigma_h\right)\boldsymbol{1}_{[0,1]^m}\label{infinitesum}\end{equation} where $\boldsymbol{\gamma}_{h}=\{\boldsymbol\eta_{h},\bSigma_h\}$ are the pre-truncation means and the pre-truncation covariances in the $h^{\mathrm{th}}$ mixture component and $c_h$ is a normalizing constant for the $h^{\mathrm{th}}$ multivariate truncated normal distribution and is a function of $\bSigma_h$ and $\boldsymbol \eta_h$. The indicator variable $\boldsymbol{1}_{[0,1]^m}$ implies truncation to the interval $[0,1]^m$. 

To facilitate Bayesian inference, approximate (\ref{infinitesum}) with the sum of the first $H$ terms. Then,

$$f_{\boldsymbol{D}}(\boldsymbol{d}\mid \{\boldsymbol{\gamma}_h\})\approx\sum_{h=1}^{H}w_h c_h \mathcal{N}\left(\boldsymbol{d}\mid \boldsymbol\eta_{h},\bSigma_h\right)\boldsymbol{1}_{[0,1]^m},$$ where $\boldsymbol{\gamma}_{h}=\{\boldsymbol\eta_{h},\bSigma_h\}$.
Next, define latent mixture component indicators $Z_i$ specifying to which of the $H$ components subject $i$ belongs.
 Then, $Z_i$ is categorically distributed meaning $$\Pr(Z_i = h)=\prod_{h=1}^H p_h^{\boldsymbol{1}(Z_i=h)} \text{ where }\sum_h p_h =1.$$

and $$f_{\boldsymbol{D}}(\boldsymbol{d}_i\mid Z_i = h,\boldsymbol{\gamma}_h )= c_h\mathcal{N}\left(\boldsymbol{d}_i\mid \boldsymbol\eta_{h},\bSigma_h\right)\boldsymbol{1}_{[0,1]^m}.$$    

The constant $H$ can be determined by running a preliminary Gibbs sampler with a large number of components. Then, observe what is the highest mixture component indicator value, and choose $H$ such that it is higher than the number of DP clusters.
We now describe how to draw the weights. Conditional on $\bZ$, for all $h \in \{1,...,H-1\}$, the posterior of the $w'_h$'s from the stick-breaking representation are:
$$f(w'_h\mid \alpha,\bZ)=\mathrm{Beta}\left(w_h'\mid 1+\sum_{i=1}^n \boldsymbol{1}(Z_i=h),\alpha+\sum_{i=1}^n \boldsymbol{1}(Z_i>h)\right).$$
Set $w'_H=1$ and the mixture weights to $w_h = w'_h \prod_{k<h}(1-w'_k)$.  Subsequently, assuming the prior $f(\alpha)$ for the concentration parameter $\alpha$, the posterior is $$f(\alpha \mid \bZ,\boldsymbol w')\propto f(\alpha)\prod_{h=1}^H \mathrm{Beta}\left(w_h'\mid 1+\sum_{i=1}^n \boldsymbol{1}(Z_i=h),\alpha+\sum_{i=1}^n \boldsymbol{1}(Z_i>h)\right).$$ By the specification of the joint distribution of potential compliances, the posterior for the mixture component indicator is: \begin{align*}\Pr(Z_i=h\mid \boldsymbol{D}_i=\boldsymbol{d}_i,\boldsymbol{\gamma}_h)&\propto f_{\boldsymbol{D}}(\boldsymbol{d}_i\mid Z_i = h,\boldsymbol{\gamma}_h)\Pr(Z_i=h)\\
    &\propto w_h c_h \mathcal{N}\left(\boldsymbol{d}_i\mid \boldsymbol{\eta}_{h},\bSigma_h\right)\boldsymbol{1}_{[0,1]^m}.
\end{align*} 
Individuals with characteristics which are most consistent with the $h^{th}$ mixture component will have $Z=h$ with high probability. This demonstrates the clustering property of the Dirichlet process mixture. Assume that the Dirichlet process base measure $G_0$ has first component which is flat and second component which is Inverse-Wishart. Then, the posterior of the pre-truncation means $\boldsymbol{\eta}_h$ are: \begin{align*}f(\boldsymbol{\eta}_{h}\mid \boldsymbol{D}_i:Z_i=h,\bSigma_h,\bZ)&\propto \prod_{i:Z_i=h} \dfrac{\mathcal{N}\left(\boldsymbol{D}_i\mid \boldsymbol{\eta}_{h},\bSigma_h\right)}{\int_{\boldsymbol 0}^{\boldsymbol 1} \mathcal{N}(\bx\mid\boldsymbol{\eta}_h,\bSigma_h)d\boldsymbol x}.
\end{align*}
 The posterior for the pre-truncation covariance $\bSigma_{h}$ are: \begin{align*}f(\bSigma_h\mid \boldsymbol{D}_i:Z_i=h,\boldsymbol\eta_{h},\bZ)&\propto \mathrm{IW}(\bSigma_h\mid m,I_m)\prod_{i:Z_i=h} \dfrac{\mathcal{N}\left(\boldsymbol{D}_i\mid \boldsymbol{\eta}_{h},\bSigma_h\right)}{\int_{\boldsymbol 0}^{\boldsymbol 1} \mathcal{N}(\bx\mid\boldsymbol{\eta}_h,\bSigma_h)d\boldsymbol x}.
\end{align*}
\section{Gibbs sampler}

Posterior inference is carried out using a Gibbs sampler. First, draw the missing potential compliances for each individual. Then, draw the parameters from the conditional posteriors at each step and iterate.
Since the posteriors of $\alpha$, $\bSigma_h$, and $\boldsymbol{\eta}_{h}$ do not have a closed form of a known distribution, Metropolis-Hastings steps are taken within the Gibbs sampler at each iteration. 

\begin{enumerate}
    \item Fix $H$, the number of DP mixture components.
    \item Draw $\boldsymbol{D}_{i}^{\mathrm{mis}},i=1,...,n$, from the conditional posterior $f(\boldsymbol{D}_{i}^{\mathrm{mis}}\mid \boldsymbol{D}_{i}^{\mathrm{obs}},Y_i,\bbeta,\sigma^2,Z_i=h,\boldsymbol{\gamma}_h)$. The expression for the posterior may be derived using facts about the conditional distribution of multivariate normal distribution. In particular, the conditional posterior density is given by the following: 
    \begin{align*}
    f(\boldsymbol{D}_{i}^{\mathrm{mis}}\mid \boldsymbol{D}_{i}^{\mathrm{obs}},Y_i,\bbeta,\sigma^2,Z_i=h,\boldsymbol{\gamma}_h)&\propto f(Y_i \mid \bD_i,\bbeta,\sigma^2,Z_i=h,\bgamma_h)f(\bD^{mis}_i\mid\bD^{obs}_i,Z_i=h,\bgamma_h)\\
    &\propto \mathcal{N}(Y_i\mid \bD_i^{\top}\bbeta,\sigma^2)\mathcal{N}(\bD^{mis}_i \mid \bD^{obs}_i,\bgamma_h,Z_i=h)\boldsymbol{1}_{[0,1]^m}.
    \end{align*}
    Note that the product of the normal density and truncated normal density will also be truncated normal and depend on $Y_i$.
    \item Draw $\boldsymbol{\eta}_{h},h=1,...,H$, from their conditional posteriors $f(\boldsymbol{\eta}_{h}\mid \boldsymbol{D}_i: Z_i=h,\boldsymbol{\gamma}_h)$ using a Metropolis-Hastings step:
    \begin{enumerate}
        \item Propose a point $\boldsymbol{\eta}_h^{\mathrm{prop}}\sim \mathcal{N}(\bar{\boldsymbol{D}}_h,n_h^{-1}\bSigma_h)$ and accept with probability
        $$\alpha_{\boldsymbol{\eta_h}}=\min\left(1,\dfrac{f(\boldsymbol{\eta}_{h}^{\mathrm{prop}}\mid \boldsymbol{D}_i: Z_i=h,\bSigma_h)f(\boldsymbol{\eta}_h^{\mathrm{current}}\mid \bar{\boldsymbol{D}}_h,n_h^{-1}\bSigma_h) }{f(\boldsymbol{\eta}_{h}^{\mathrm{current}}\mid \boldsymbol{D}_i: Z_i=h,\bSigma_h)f(\boldsymbol{\eta}_h^{\mathrm{prop}}\mid \bar{\boldsymbol{D}}_h,n_h^{-1}\bSigma_h)}\right).$$
    \end{enumerate}
    \item Draw $\bSigma_h,h=1,...,H$, from their conditional posterior $f(\bSigma_h\mid \boldsymbol{D}_i:Z_i=h,\boldsymbol{\eta}_h)$ using a Metropolis-Hastings step:
    \begin{enumerate}
        \item Propose a point $\bSigma_h^{\mathrm{prop}}\sim \mathrm{Wishart}(1000,\bSigma_h^{\mathrm{current}}/1000)$ and accept with probability:
        $$\alpha_{\bSigma_h}=\min\left(1,\dfrac{f(\bSigma_{h}^{\mathrm{prop}}\mid \boldsymbol{D}_i:Z_i=h,\boldsymbol{\eta}_{h})f_{\mathrm{Wishart}}({\bSigma_h^{\mathrm{current}}}\mid \bSigma^{\mathrm{prop}}_h)}{f(\bSigma_h^{\mathrm{current}}\mid \boldsymbol{D}_i:Z_i=h,\boldsymbol{\eta}_{h})f_{\mathrm{Wishart}}(\bSigma_h^{\mathrm{prop}}\mid\bSigma^{\mathrm{current}}_h)}\right).$$
    \end{enumerate}
    \item Given $\boldsymbol{w}$ and $\boldsymbol{D}_i$, draw $Z_i$, $i=1,...,n$, from a categorical distribution with probabilities $$\Pr(Z_i=h\mid w_h,\boldsymbol{D}_i,\boldsymbol{\eta}_h,\bSigma_h)\propto w_h c_h \mathcal{N}(\boldsymbol{D}_i\mid \boldsymbol{\eta}_h,\bSigma_h)\boldsymbol{1}_{[0,1]^m}.$$
    \item Set $w'_H=1$ and
    \begin{enumerate}
        \item Draw $w'_h, h=1,...,H-1$, from $\mathrm{Beta}\left(1+\sum_{i=1}^n \boldsymbol{1}(Z_i=h),\alpha+\sum_{i=1}^n\boldsymbol{1}(Z_i>h)\right).$
        \item Update $w_h=w'_h\prod_{k<h}(1-w'_k)$.
    \end{enumerate}
    \item Draw $\alpha$ from the following using a Metropolis-Hastings step $$f(\alpha\mid \boldsymbol{w}',\boldsymbol Z)\propto f(\alpha)\prod_{h=1}^H \mathrm{Beta}\left(w_h'\mid 1+\sum_{i=1}^n \boldsymbol{1}(Z_i=h),\alpha+\sum_{i=1}^n\boldsymbol{1}(Z_i>h)\right).$$
    \begin{enumerate}
        \item Propose $\alpha^{\mathrm{prop}}$ from Gamma$(1,1)$ and accept with probability 
        $$\alpha_{\alpha}=\dfrac{\displaystyle\prod_{h=1}^{H-1}\mathrm{Beta}\left(w_h'\mid 1+\sum_{i=1}^n \boldsymbol{1}(Z_i=h),\alpha^{\mathrm{prop}}+\sum_{i=1}^n\boldsymbol{1}(Z_i>h)\right)}{\displaystyle\prod_{h=1}^{H-1}\mathrm{Beta}\left(w'_h \mid 1+\sum_{i=1}^n \boldsymbol{1}(Z_i=h),\alpha^{\mathrm{current}}+\sum_{i=1}^n\boldsymbol{1}(Z_i>h)\right)}.$$
    \end{enumerate}
    \item Draw $\bbeta^{(1)},\bbeta^{(2)},...,\bbeta^{(K)}$  from a Bayesian linear regression with a flat prior imposing the equality constraints and draw the corresponding residual variances $\sigma_{1Y}^2,...,\sigma_{KY}^2$ where $K$ is the number of treatment sequences.
       The posterior for $\bbeta^{(k)}$ is given by $\mathcal{N}(\hat{\bbeta}^{(k)},\sigma_k^2(\bX_k^{\top}\bX_k)^{-1})$ where $\bX_k$ is the design matrix and $\hat{\bbeta}^{(k)}=(\bX_k^{\top}\bX_k)^{-1}\bX^{\top}_k\boldsymbol{Y}_k$. The conditional posterior will also be normal.     The posterior for $\sigma^2_{kY}$ is given by 
        $\mathrm{Inv}-\chi^2(s^2,n_k-J)$ where $s^2 = (\bY_k-\bX_k\hat{\bbeta}^{(k)})^{\top}(\bY_k-\bX_k\hat{\bbeta}^{(k)})/(n_k-J)$ where $J$ is the dimension of $\hat{\bbeta}^{(k)}$.
    \item Iterate until convergence.    
\end{enumerate}

\section{Regression model and assumptions for outcomes}
\subsection{Assumptions}

We make the following assumption for the design of the SMART to reduce the number of relevant potential compliances among responders and non-responders: 
\begin{itemize}
\item[A.1]{\it No cross-treatment access}. $\mathrm{Pr}(D_{2}^{R} =0 \mid S=0)=\mathrm{Pr}(D_{2}^{NR} =0 \mid S=1)=1$.
\end{itemize}

Assumption A.1 implies that individuals with a certain response status do not have access to the treatment options under the other response status. This is a typical assumption in compliance literature that often holds in clinical trials. 
Under A.1, the potential outcome of treatment sequences for responders and non-responders are by design independent of potential compliances for non-responders $(D_{21}^{\mathrm{NR}},D_{22}^{\mathrm{NR}})$ and responders $(D_{21}^{\mathrm{R}},D_{22}^{\mathrm{R}})$, respectively. That is $\E[Y \mid \boldsymbol{D}, S=1] = \E[Y \mid D_{11},D_{12},D_{21}^{\mathrm{R}},D_{22}^{\mathrm{R}},S=1]$, and $\E[Y \mid \boldsymbol{D}, S=0] = \E[Y \mid D_{11},D_{12},D_{21}^{\mathrm{NR}},D_{22}^{\mathrm{NR}},S=0]$. 
In ENGAGE, subjects who respond to either MI-IOP or MI-PC at stage 1 (i.e., engagers) could not have received the non-responder treatments. Therefore, we model responders using only the stage-1 potential compliances $(D_{11},D_{12})$ (i.e., $\E[Y \mid \boldsymbol{D}, S=1] = \E[Y \mid D_{11},D_{12},S=1]$). In addition, we assume the following:

\begin{itemize}
\item[A.2]{\it Identifiability}. For each latent compliance value in the $k^{th}$ treatment sequence, there is a treatment sequence in which the corresponding compliance value is observed and its effect is identifiable. 
\item[A.3] {\it Homogeneity.} Suppose $D_{nm}$ is observed in treatment sequence $k$ and unobserved in sequence $k'$, we assume $\delta(\boldsymbol{D} \setminus D_{nm})=\E[Y_k \mid \boldsymbol{D},D_{nm}=d] - \E[Y_{k'} \mid \boldsymbol{D},D_{nm}=d]$ does not depend on $D_{nm}$. The effect of the compliance is homogeneous at least across two treatment sequences where one treatment sequence involves the observed compliance value and other involves the latent compliance value.  
\item[A.4] {\it Homogeneity for Stage-1 Response.} Suppose $D_{nm}$ is observed in treatment sequence $k$ and unobserved in sequence $k'$, we assume $\delta_S\{\boldsymbol{D} \setminus (D_{11},D_{12})\}=\mathrm{logit}\Pr(S=1 \mid A_1=1, D_{11},D_{12}) - \mathrm{logit}\Pr(S=1 \mid A_1=-1,D_{11},D_{12})$ does not depend on $D_{11},D_{12}$. The effect of the compliance is homogeneous across groups randomized to $A_1=1$ and $A_1=-1$.
\item[A.5] {\it Ignorability of opposite branch compliance} $S_i \indep D_{12i}\mid A_{1i}=+1,D_{11i},\bX$ and $S_i \indep D_{11i}\mid A_{1i}=-1,D_{12i},\bX$ where $\bX$ are baseline covariates.
\item[A.6] {\it No direct effect of stage-1 treatment on stage-2 potential compliance} $D_{2,A_1,A_2,S}=D_{2,A_2,S}$.
\end{itemize}

 The compliance classes are partially latent. That is, there is no individual in the data for whom all the compliance values are observed. This implies that the joint density of the potential compliances that is required in our Bayesian procedure is not identifiable in the absence of assumptions A.2 and A.3. Moreover, in the conditional outcome models, the effect of unobserved compliance value is not identifiable. For example, in ENGAGE, if we consider $\E[Y_1 \mid D_{11},D_{12}] = \beta_0^{(1)} + \beta_1^{(1)} D_{11}+ \beta_2^{(1)} D_{12}$, the parameters cannot be identified because $D_{12}$ is unobserved which also affects the convergence of the MCMC algorithm through the augmentation step (i.e., step 2 in the Gibbs sampler). In fact, estimating the parameters in the outcome model and augmenting the missing compliance values are tied together. In combination, assumptions A.2 and A.3 close this loop by equating the parameter value corresponding to a missing compliance in the outcome model for the $k^{th}$ treatment sequence to the corresponding parameter value in the outcome model for another treatment sequence in which that compliance value is observed. Although not explicitly stated, \cite{schwartz2011bayesian} also imposed these two identifiability assumptions in a two-arm randomized trial setting. The following theorem formalizes the identifiability result. 

\begin{theorem}
Assume A.2 and A.3 hold. Then, the regression coefficients for both observed and latent potential compliances corresponding to the $l^{th}$ treatment sequence, $\beta_j^{(l)}$, are identified.
\end{theorem}

\subsection{ENGAGE modeling}
We now specifically discuss modeling for the ENGAGE SMART. In this study, assumption A.1 is satisfied by design. The key is to impose certain independence assumptions such that assumptions A.2 and A.3 are satisfied. In ENGAGE, there are 3 potential compliances $(D_{11},D_{12}, D_{22})$. Thus a linear model including all the main potential compliance effects are
  \begin{align*}
Y_k &= \beta_0^{(k)}+\beta_1^{(k)} D_{11} + \beta_2^{(k)} D_{12 }+\epsilon_{k}, \hspace{.3in} k = 1,4 \\
Y_{k}&=\beta_0^{(k)}+\beta_1^{(k)} D_{11} + \beta_2^{(k)} D_{12 }+ \beta_3^{(k)} D_{22 }+\epsilon_k, \hspace{.3in} k=2,3,5,6,
\end{align*} 

where $k\in\{1,4\}$ and $k\in\{2,3,5,6\}$ correspond to treatment sequences for responders and non-responders, respectively and $\epsilon_k \sim \mathcal{N}(0,\sigma^2_k)$. For responders, conditional independence between the potential outcome and the stage-2 treatment potential compliance to which the subject was not assigned, $D_{22}$ (MI-PC) holds by design (i.e., $Y_{ik}\indep D_{22i}\mid \{D_{11i},D_{12i}\}$, $k=1,4$).
These are very general models that allow the intercepts and main effects for each compliance values to depend on $k$. However, some of the parameters in these models are not identifiable. For example, for $k=1$, the compliance value $D_{12}$ is unobserved and because the effect of $D_{12}$ is heterogeneous (i.e., A.2 and A.3 are violated), the parameter $\beta_{2}^{(1)}$ is not identifiable.  To make the parameters identifiable we impose some restrictions on the parametrization of these models. 

Assume that the outcome model for the $1^{st}$ treatment sequence is conditionally independent of $D_{12}$ given $D_{11}$ (i.e., $Y_{i1}\indep D_{12i}\mid D_{11i}$). 
In addition, for non-responders to their stage-1 treatment, assume conditional independence between the potential outcome and the opposite branch stage-1 treatment potential compliance. Formally,  
\begin{align*}
Y_{ik}\indep D_{12i}\mid D_{11i}, k = 2,3  \hspace{.3in}  \text{ and } \hspace{.3in} Y_{ik}\indep D_{11i}\mid D_{12i}, k=5,6.
\end{align*} 

We also impose certain effect homogeneity conditions in order to identify the regression coefficients in the outcome-potential compliance model. We set $\beta_0^{(4)}=\beta_0^{(1)}$ which assumes that in the absence of any intervention, treatment sequences 1 and 4 have the same average potential outcome. This is reasonable because the only distinguishing feature for such subjects would be to which group they were randomized initially. Similarly, we assume $\beta_0^{(3)}=\beta_0^{(2)}$ and $\beta_0^{(6)}=\beta_0^{(5)}$. We also impose the following compliance effect homogeneity assumptions:  $\beta_1^{(4)}=\beta_1^{(1)}$; $\beta_3^{(3)}=\beta_3^{(2)}$ and $\beta_3^{(6)}=\beta_3^{(5)}$. The former implies that the rate of change with respect to potential compliance for the outcome in treatment sequence 4 (responders to initial treatment) as a function of MI-IOP potential compliance is the same as that in the opposite SMART branch. The second homogeneity assumption $\beta_3^{(3)}=\beta_3^{(2)}$ implies that the non-responders to stage-1 treatment MI-IOP for the potential compliance corresponding to stage-2 MI-PC have a rate of change which is the same between groups. The interpretation of $\beta_3^{(6)}=\beta_3^{(5)}$ is similar.
\begin{table}[t]
\centering
\begin{tabular}[t]{cccc}
\toprule
 k& Treatment Sequence & \makecell{Observed Potential\\ Compliances} & Latent Potential Compliance\\
\toprule
1 & $A_1=+1,S=1$ & \makecell{$D_{11}$}  & NA\\
\midrule
2 & $A_1=+1,S=0,A^{NR}_2=+1$ & \makecell{$D_{11}, D_{22}$}  & NA\\
\midrule
3 & $A_1=+1,S=0,A^{NR}_2=-1$&  $D_{11}$  & $D_{22}$ \\
\midrule
4&$A_1=-1,S=1$&$D_{12}$  & $D_{11}$ \\
\midrule
5&$A_1=-1,S=0,A^{NR}_2=+1$ & $D_{12}$, $D_{22}$  &NA\\
\midrule
6&$A_1=-1,S=0,A^{NR}_2=-1$ &$D_{12}$ & $D_{22}$ \\
\bottomrule
\end{tabular}
\caption{Potential compliances modeled in each treatment sequence of the ENGAGE SMART.}
\label{tab:EDTRsModelledENGAGE}
\end{table}

Then, the potential outcome models for responders to stage-1 treatment are 
\begin{align*}
    Y_{i1}&=\beta_0^{(1)}+\beta_1^{(1)}D_{11i}+\epsilon_{1i}\\
    Y_{i4}&=\beta_0^{(4)}+\beta_1^{(4)}D_{11i}+\beta_2^{(4)}D_{12i}+\epsilon_{4i}
\end{align*}
where $\epsilon_{ki}\sim \mathcal{N}(0,\sigma^2_k), k=1,4$.
So, treatment sequence 1 has potential outcome which is a function of MI-IOP potential compliance and sequence 4 has potential outcome which is also a function of stage-1 MI-PC potential compliance.

 The treatment sequences for non-engagers are \begin{align*}
    Y_{ik}&=\beta_0^{(k)}+\beta_1^{(k)}D_{11i}+\beta_3^{(k)}D_{22i}+\epsilon_{ki}, \hspace{.3in} k=2,3\text{ and }\\Y_{ik}&=\beta_0^{(k)}+\beta_2^{(k)}D_{12i}+\beta_3^{(k)}D_{22i}+\epsilon_{ki},  \hspace{.3in} k=5,6.
\end{align*}
where $\epsilon_{ki}\sim \mathcal{N}(0,\sigma^2_k),k=2,3,5,6$.
Treatment sequences 2 and 3 are functions of the potential stage-1 MI-IOP compliance and the potential stage-2 MI-PC compliance. Furthermore, treatment sequences 5 and 6 are functions of the potential stage-1 MI-PC compliance and the potential stage-2 MI-PC compliance.

In combination the imposed restrictions lead to parameterizations of the outcome models that satisfy assumptions A.2 and A.3. For example, $\beta_1^{(4)}$ is identifiable because $\beta_1^{(1)}$ is identifiable (A.2 is satisfied) and, for $D_{11}$, $\delta(\boldsymbol{D} \setminus D_{11})=\E[Y^1 \mid \boldsymbol{D},D_{11}=d] - \E[Y^{4} \mid \boldsymbol{D},D_{11}=d]$ does not depend on $D_{11}$ (i.e., A.3 is satisfied). The identification of the other parameters follows similarly. 

\subsection{General modeling}

Recall from Section 3 that 5 potential compliances are being modelled ($D_{11},D_{12},D_{22}^{\mathrm{R}}$, $D_{21}^{\mathrm{NR}},D_{22}^{\mathrm{NR}}$). Therefore, models for each treatment sequence potential outcomes would be 
 \begin{align*}
Y_k &= \beta_0^{(k)}+\beta_1^{(k)} D_{11} + \beta_5^{(k)} D_{12}+\beta_2^{(k)} D_{22}^{\mathrm{R}}+\epsilon_{k}, \hspace{.3in} k = 1,2,5,6 \\
Y_{k}&=\beta_0^{(k)}+\beta_1^{(k)} D_{11} + \beta_5^{(k)} D_{12 }+ \beta_3^{(k)} D_{21 }^{\mathrm{NR}}+\beta_4^{(k)}D_{22}^{\mathrm{NR}}+\epsilon_k, \hspace{.3in} k=3,4,7,8,
\end{align*} 
where $k \in \{1,2,5,6\}$ and $k \in \{3,4,7,8\}$ correspond to treatment sequences for responders and non-responders, respectively and $\epsilon_{ki}\sim \mathcal{N}(0,\sigma^2_k), k=1,...,8$.

Assumption A.1 holds by design just as in ENGAGE. We impose independence assumptions so that A.2 and A.3 hold. For responders, we have conditional independence between the potential outcomes and stage-2 potential compliance corresponding to treatments the subject was not assigned $D_{21}^{\mathrm{NR}},D_{22}^{\mathrm{NR}}$ by design:
\begin{align*}
 Y_{i1}&\indep \{D_{21i}^{\mathrm{NR}},D_{22i}^{\mathrm{NR}}\}\mid D_{11i}&Y_{i5}&\indep \{D_{21i}^{\mathrm{NR}},D_{22i}^{\mathrm{NR}}\}\mid D_{12i}\\ 
 Y_{i2}&\indep \{D_{21i}^{\mathrm{NR}},D_{22i}^{\mathrm{NR}}\}\mid D_{11i},D_{22i}^{\mathrm{R}}&Y_{i6}&\indep \{D_{21i}^{\mathrm{NR}},D_{22i}^{\mathrm{NR}}\}\mid D_{12i},D_{22i}^{\mathrm{R}}
\end{align*}

Similarly, for non-responders, $D_{22i}^{\mathrm{R}}$:
\begin{align*}
Y_{i3}&\indep D_{22i}^{\mathrm{R}}\mid D_{11i},D_{21i}^{\mathrm{NR}}&Y_{i7}&\indep D_{22i}^{\mathrm{R}}\mid D_{12i},D_{22i}^{\mathrm{NR}}\\ Y_{i4}&\indep D_{22i}^{\mathrm{R}}\mid D_{11i},D_{21i}^{\mathrm{NR}}, D_{22i}^{\mathrm{NR}}&Y_{i8}&\indep D_{22i}^{\mathrm{R}}\mid D_{12i},D_{21i}^{\mathrm{NR}},D_{22i}^{\mathrm{NR}}
\end{align*}

Without imposing identifiability constraints, the assumptions A.2 and A.3 will be violated and regression coefficients will not be identifiable.
Assume the following conditional independence conditions between the potential outcomes and the potential compliances:
\begin{align*}
    Y_{i1} &\indep \{D_{22i}^{\mathrm{R}},D_{12i}\}\mid D_{11i}&Y_{i5} &\indep \{D_{11i},D_{21i}^{\mathrm{NR}},D_{22i}^{\mathrm{NR}}\}\mid D_{12i}\\
    Y_{i2} &\indep D_{12i}\mid D_{11i},D_{22i}^{\mathrm{R}}&Y_{i6} &\indep D_{11i}\mid D_{12i},D_{22i}^{\mathrm{R}}\\
    Y_{i3} & \indep \{D_{22i}^{\mathrm{NR}},D_{12i}\}\mid D_{11i},D_{21i}^{\mathrm{NR}}&Y_{i7} & \indep \{D_{11i},D_{22i}^{\mathrm{NR}}\}\mid D_{12i},D_{21i}^{\mathrm{NR}}\\
    Y_{i4} & \indep D_{12i}\mid D_{11i},D_{21i}^{\mathrm{NR}},D_{22i}^{\mathrm{NR}}&Y_{i8} & \indep D_{11i}\mid D_{12i},D_{21i}^{\mathrm{NR}},D_{22i}^{\mathrm{NR}}.
\end{align*}

We impose the following homogeneity constraints in order to identify the regression coefficients: we set $\beta_0^{(4)}=\beta_0^{(3)}$ which assumes in the absence of any intervention, treatment sequences $3$ and $4$ have the same average potential outcome. This is reasonable because the only distinguishing feature for such subjects would be to which group they were randomized intially. Similarly, $\beta_0^{(8)}=\beta_0^{(7)}$. We also impose the following compliance effect homogeneity assumptions: 
$\beta_1^{(4)}=\beta_1^{(3)}$; $\beta_5^{(8)}=\beta_5^{(7)}$; $\beta_3^{(4)}=\beta_3^{(3)}$ and $\beta_3^{(8)}=\beta_3^{(7)}$. These consrtaints imply that the rate of change with respect to a particular potential compliance for a given treatment sequence as a function of the potential compliance is the same as in another treatment sequence. The identifiability constraints guarantee that A.2 and A.3 hold.

Our assumptions imply the following potential outcome models: \begin{align*}
    Y_1 & = \beta_0^{(1)}+\beta_1^{(1)}D_{11}+\epsilon_1&Y_5 & = \beta_0^{(5)}+\beta_5^{(5)}D_{12}+\epsilon_5\\
    Y_2 & = \beta_0^{(2)}+\beta_1^{(2)}D_{11}+\beta_2^{(2)}D_{22}^{\mathrm{R}}+\epsilon_2&Y_6 & = \beta_0^{(6)}+\beta_5^{(6)}D_{12}+\beta_2^{(6)}D_{22}^{\mathrm{R}}+\epsilon_6\\
    Y_3 & = \beta_0^{(3)}+\beta_1^{(3)}D_{11}+\beta_3^{(3)}D_{21}^{\mathrm{NR}}+\epsilon_3&Y_7 & = \beta_0^{(7)}+\beta_5^{(7)}D_{12}+\beta_3^{(7)}D_{21}^{\mathrm{NR}}+\epsilon_7\\
    Y_4 & = \beta_0^{(4)}+\beta_1^{(4)}D_{11}+\beta_3^{(4)}D_{21}^{\mathrm{NR}}+\beta_4^{(4)}D_{22}^{\mathrm{NR}}+\epsilon_4 &Y_8 & = \beta_0^{(8)}+\beta_5^{(8)}D_{12}+\beta_3^{(8)}D_{21}^{\mathrm{NR}}+\beta_4^{(8)}D_{22}^{\mathrm{NR}}+\epsilon_8
\end{align*}
where $\epsilon_{ki}\sim \mathcal{N}(0,\sigma^2_k), k=1,...,8$.
Assumptions A.2 and A.3 are satisfied, so by Theorem 1, the coefficients in the above models are identified.
We have described how to model $\E[Y\mid A_{1}=a_{1},S,A_2=a_2,\bD]$. Next, we describe how to model $\Pr(S=1\mid A_1=a_1,\bD)$ so that we can estimate the principal causal effects.

\begin{table}[t]
\centering
\begin{tabular}[t]{cccc}
\toprule
 k& Treatment Sequence & \makecell{Observed Potential\\ Compliances} & Latent Potential Compliance\\
\toprule
1 & $A_1=+1,S=1,A_2^{\mathrm{R}}=+1$ & \makecell{$D_{11}$}  & NA\\
\midrule
2 & $A_1=+1,S=1,A^{\mathrm{R}}_2=-1$ & \makecell{$D_{11}, D_{22}^{\mathrm{R}}$}  & NA\\
\midrule
3 & $A_1=+1,S=0,A^{\mathrm{NR}}_2=+1$&  $D_{11},D_{21}^{\mathrm{NR}}$  & NA \\
\midrule
4 & $A_1=+1,S=0,A^{\mathrm{NR}}_2=-1$&  $D_{11},D_{22}^{\mathrm{NR}}$  & $D_{21}^{\mathrm{NR}}$ \\
\midrule
5&$A_1=-1,S=1,A^{\mathrm{R}}_2=+1$&$D_{12}$  & NA \\
\midrule
6&$A_1=-1,S=1,A^{\mathrm{R}}_2=-1$ & $D_{12},D_{22}^{\mathrm{R}}$  &NA\\
\midrule
7&$A_1=-1,S=0,A^{\mathrm{NR}}_2=+1$ & $D_{12},D_{21}^{\mathrm{NR}}$ & NA\\
\midrule
8&$A_1=-1,S=0,A^{\mathrm{NR}}_2=-1$ & $D_{12},D_{22}^{\mathrm{NR}}$  &$D_{21}^{\mathrm{NR}}$\\
\bottomrule
\end{tabular}
\caption{Potential compliances modeled in each treatment sequence of the General SMART.}
\label{tab:EDTRsModelledgeneral}
\end{table}

\subsection{Stage-1 response indicator model}
We assume A.4 holds. We model the probability of response in each branch using Bayesian logistic regression with a flat prior on $\boldsymbol{\gamma}$: $$\Pr(S=1\mid A_1=a_{1j}, D_{11},D_{12})=\expit\{(1,D_{11},D_{12},a_{1j})^{\top}\boldsymbol{\gamma}\}$$
where $\expit(\cdot)=\dfrac{\exp(\cdot)}{1+\exp(\cdot)}$. In particular, the posterior of $\bgamma$ is proportional to the following:

$$\prod_{i=1}^n\left[\expit\{(1,D_{11i},D_{12i},a_{1ji})^{\top}\boldsymbol{\gamma}\}\right]^{S_i}\left[1-\expit\{(1,D_{11i},D_{12i},a_{1ji})^{\top}\boldsymbol{\gamma}\}\right]^{1-S_i}$$

We draw from the posterior using Metropolis-Hastings (\citealp{hastings1970monte}). We simulate 10000 draws and consider every 10th draw. Under assumption A.4, the parameter $\gamma$ can be identified.  
Alternatively, one could assume A.5 in which case the model would omit one of the $D_{1j}$'s and model each treatment sequence $A_1=-1$ and $A_1=+1$ separately.

\section{Model selection}

We perform model selection using the Watanabe–Akaike information criterion \citep{gelman2013bayesian}. This information criterion aims to estimate the expected log pointwise predictive density (elppd) for a new data set with $n$ points which is given by \begin{align*}\mathrm{elppd}&=\sum_{i=1}^n \E_{\tilde{y}_i}\left[\log p(\tilde{y}_i\mid\boldsymbol y)\right]\\
&=\sum_{i=1}^n \E_{\tilde{y}_i}\left[\log \int p(\tilde{y}_i\mid\theta)p(\theta\mid \boldsymbol{y})d\theta\right]\\&=\sum_{i=1}^n\int\left[ \log \int p(\tilde{y}_i\mid \theta)p(\theta\mid \boldsymbol{y})d\theta \right]p(\tilde{y}_i)d\tilde{y}_i
\end{align*}
where $p(\tilde{y}_i)$ is the marginal distribution of each data point and $\boldsymbol y$ is the observed data. 
Watanabe–Akaike information criterion averages over the posterior distribution instead of conditioning on, for example, the posterior mean. It is estimating the average log predictive density for a new data set as opposed to just a single point. It is a fully Bayesian information criterion in that it uses the posterior distribution.

We treat the imputed potential compliances as parameters. 
The Watanabe–Akaike information criterion assesses the predictive accuracy of the model applied to new data given the observed data and penalizes for model complexity as a bias correction.
An estimate of the information criterion (WAIC) for treatment sequence $j$ is given as follows  (\citealp{gelman2013bayesian}):
\begin{align*} 
\displaystyle\widehat{\mathrm{WAIC}}&=-2\sum_{i=1}^{n_j}\log\left\{\dfrac{1}{K}\sum_{k=1}^K f(Y_i\mid \bD_{obs},\bD_{mis}^{(k)},\bbeta_j^{(k)},\sigma_j^{2, (k)})\right\}\\&+2\sum_{i=1}^{n_j}\widehat{\var}_{\bD_{mis},\bbeta_j,\sigma_j^2}\left[\log\left\{f\left(Y_i\mid \boldsymbol D_{obs},\boldsymbol{D}_{mis},\bbeta_j,\sigma_j^2\right)\right\}\right].
\end{align*}

 The expression has a factor of $-2$ so that it is on the deviance scale.
 Lower values of the information criterion indicate better model fit.

\section{Multiple comparisons with the best}
Our goal is to identify the embedded DTRs with the greatest efficacy incorporating compliance class. Multiple comparisons with the best (MCB) provides clinicians with a way to interpret compliance classes' impact on outcomes. It provides inference for the optimal embedded DTRs under varying levels of compliance. MCB was leveraged in the SMART setting in \cite{ertefaie2015identifying,artman2018power}. MCB was based off a normal parametric model in the frequentist setting. Below, we describe an extension to the Bayesian setting which is calculated using draws from the posterior.

MCB entails constructing simultaneous one-sided upper credible limits for the difference between each embedded DTR and the optimal embedded DTR. Here, the Bonferroni correction was used to construct 95\% credible intervals which control the false positive rate. When choosing an embedded DTR from the set of best, it makes sense to also consider side-effect profiles. With MCB, one may offer a set of optimal embedded DTRs which are not statistically distinguishable from the best for the given data. Consequently, it allows the clinician to choose an optimal embedded DTR which takes into account cost and side-effect burden known externally from the study. We apply the method to four different compliance classes as well as the intention-to-treat group. If the interval covers zero, it is statistically indistinguishable from the best. If it does not cover zero, it is excluded from the set of best (optimal embedded DTRs, $\hat{\mathcal{B}}$ below).

 \begin{table*}[t]
\centering
\begin{tabular*}{420pt}{@{\extracolsep\fill}cccccccc@{\fill}}
\toprule
&\multicolumn{6}{@{}c@{}}{Bias (Standard Errors) for Main Effects Only Model}\\
\toprule
&\multicolumn{6}{@{}c@{}}{Correlation = 0.2}\\
\toprule
  Par.& Trt. Seq. 1& Trt. Seq. 2 & Trt. Seq. 3& Trt. Seq. 4&Trt. Seq. 5&Trt. Seq. 6\\
\toprule
$\beta_0$ & 0.00 (0.05) & 0.01 (0.06) & 0.01 (0.06) & 0.00 (0.05) & 0.00 (0.05) & 0.00 (0.05)\\
$\beta_1$ & 0.00 (0.08) & 0.01 (0.09) & 0.01 (0.13) & 0.00 (0.08) & -- (--) & -- (--)\\
$\beta_2$ & -- (--) & -- (--) & -- (--) & 0.04 (0.07) & 0.00 (0.07) & 0.00 (0.10)\\
$\beta_3$ & -- (--) & 0.00 (0.08) & 0.00 (0.08) & -- (--) & 0.01 (0.07) & 0.01 (0.07)\\
\bottomrule
\end{tabular*}
\centering
\begin{tabular*}{420pt}{@{\extracolsep\fill}cccccccc@{\fill}}
&\multicolumn{6}{@{}c@{}}{Correlation = 0.5}\\
\toprule
  Par.& Trt. Seq. 1& Trt. Seq. 2 & Trt. Seq. 3& Trt. Seq. 4&Trt. Seq. 5&Trt. Seq. 6\\
\toprule
$\beta_0$ & 0.00 (0.05) & 0.01 (0.06) & 0.01 (0.06) & 0.00 (0.05) & 0.00 (0.05) & 0.00 (0.05)\\
$\beta_1$ & 0.00 (0.08) & 0.01 (0.10) & 0.01 (0.14) & 0.00 (0.08) & -- (--) & -- (--)\\
$\beta_2$ & -- (--) & -- (--) & -- (--) & 0.07 (0.08) & 0.01 (0.07) & 0.01 (0.10)\\
$\beta_3$ & -- (--) & 0.00 (0.09) & 0.00 (0.09) & -- (--) & 0.01 (0.09) & 0.01 (0.09)\\
\bottomrule
\end{tabular*}
\centering
\begin{tabular*}{420pt}{@{\extracolsep\fill}cccccccc@{\fill}}
&\multicolumn{6}{@{}c@{}}{Correlation = 0.8}\\
\toprule
  Par.& Trt. Seq. 1& Trt. Seq. 2 & Trt. Seq. 3& Trt. Seq. 4&Trt. Seq. 5&Trt. Seq. 6\\
\toprule
$\beta_0$ & 0.00 (0.05) & 0.01 (0.06) & 0.01 (0.06) & 0.00 (0.05) & 0.00 (0.05) & 0.00 (0.05)\\
$\beta_1$ & 0.00 (0.08) & 0.01 (0.14) & 0.04 (0.19) & 0.00 (0.08) & -- (--) & -- (--)\\
$\beta_2$ & -- (--) & -- (--) & -- (--) & 0.10 (0.08) & 0.01 (0.10) & 0.04 (0.12)\\
$\beta_3$ & -- (--) & 0.00 (0.13) & 0.00 (0.13) & -- (--) & 0.01 (0.12) & 0.01 (0.12)\\
\bottomrule
\end{tabular*}
\caption{ ENGAGE simulation results for main effects only model with $n=250$ sample size and approximately $200$ replicates. $\beta_0$ is the intercept; $\beta_1$ is the coefficient corresponding to $D_{11}$ (stage-1 MI-IOP); $\beta_2$ is the coefficient corresponding to $D_{12}$ (stage-1 MI-PC); and $\beta_3$ is the coefficient corresponding to $D_{22}$ (stage-2 MI-PC). -- = not part of model.}

\label{tab:ENGAGESimResultsMainEffects}
\end{table*}
The goal of MCB is to construct a set of optimal embedded DTRs as a function of compliance class. The set of best is defined as follows: $$\hat{\mathcal{B}}=\{\mathrm{EDTR}_l\mid \mathrm{EDTR}_l\text{ is not statistically distinguishable from the best EDTR at level }\alpha \}. $$

To construct such a set, we compare each embedded DTR to the best embedded DTR by constructing one-sided upper credible intervals adjusting for multiplicity. In particular, simultaneous credible intervals for $Y^{(l)}-\max_{l'} Y^{(l')}$. Let $U_l$ denote the upper limit of the $l^{th}$ credible interval. Then, $\hat{\mathcal{B}}=\{\mathrm{EDTR}_l\mid U_l\geq 0\}.$

The embedded DTRs not contained in $\hat{\mathcal{B}}$ are significantly inferior to the optimal embedded DTR. An advantage of MCB is that it entails only $L-1$ comparisons where $L$ is the number of embedded DTRs. This may be substantially less than all pairwise comparisons.

We construct $U_1,...,U_{L-1}$ from the MCMC draws of $Y^{(l)}, l = 1,...,L$ by taking the $1-\alpha/(L-1)$ quantile of each of the draws minus the best embedded DTR outcome draw. This controls the type I error rate of excluding the true best embedded DTR. This can be done for each compliance class $\mathcal{C}_b$ yielding $\hat{\mathcal{B}}(\mathcal{C}_1),\hat{\mathcal{B}}(\mathcal{C}_2),...,\hat{\mathcal{B}}(\mathcal{C}_B)$. The embedded DTRs which are included in all plausible compliance class are a good choice for patients whose potential compliance class is unknown. Given information about what compliance class a patient will likely belong to determines which set of best embedded DTRs is optimal and subsequently, which embedded DTRs are optimal.

\section{Simulation study}

We simulated two SMARTs with the same structure as ENGAGE and the General SMART design, respectively with a sample size of $n=250$.

\subsection{ENGAGE simulation}
 Here, we summarize the ENGAGE SMART simulation.

 The potential compliances were generated from a Gaussian copula model using the \texttt{R} package \texttt{copula}. This model is fully specified by the marginal distributions and the correlation matrix of the potential compliances. We use Beta distributed margins such that the parameters for the stage-1 compliance are $(3,2)$, $(2,1)$, and the stage-2 compliance is $(2,3)$. The correlation matrix was $$\boldsymbol R=\begin{pmatrix}1&\rho&\rho\\\rho&1&\rho\\\rho&\rho&1\end{pmatrix}$$
for $\rho = 0.2,0.5,0.8$. 
We generated stage-1 response indicators as $S_i \sim \mathrm{Bern}\left\{\dfrac{\exp(D_{11}-1)}{1+\exp(D_{11}-1)}\right\}$  when  $A_{1i}=+1$  and $S_i \sim \mathrm{Bern}\left\{\dfrac{\exp(D_{12}-1.5)}{1+\exp(D_{12}-1.5)}\right\}$ when $A_{1i}=-1$.
The following potential outcome models were generated for each of the 6 treatment sequences embedded in the SMART.
\begin{align*}
    Y_{i1} &=0.7+0.6D_{11i}+\epsilon_{1i}&Y_{i4} &=0.7+0.6D_{11i}+0.6D_{12i}+\epsilon_{4i}\\
    Y_{i2}&=0.2+0.7D_{11i}+0.9D_{22i}+\epsilon_{2i}&Y_{i5} &=0.3+0.6 D_{12i}+0.7D_{22i}+\epsilon_{5i}\\
    Y_{i3}&=0.2+0.6D_{11i}+0.9D_{22i}+\epsilon_{3i}&Y_{i6} &=0.3+0.6D_{12i}+0.7D_{22i}+\epsilon_{6i}
\end{align*}
where $\epsilon_{ki}\overset{iid}{\sim} \mathcal{N}(0,0.1^2)$.
The embedded DTR regression coefficient results are summarized in Table \ref{tab:ENGAGESimResultsMainEffects}. In particular, note that the estimates are nearly unbiased across correlations. Furthermore, the standard errors are small even for the high correlation scenario. The bias and standard errors are slightly higher, but still reasonable when the correlation between potential compliance is 0.8.

\begin{table*}[t]
\centering
\begin{tabular*}{420pt}{@{\extracolsep\fill}cccccccc@{\fill}}
\toprule
&\multicolumn{6}{@{}c@{}}{Bias (Standard Errors) for Interaction Model}\\
\toprule
&\multicolumn{6}{@{}c@{}}{Correlation = 0.2}\\
\toprule
  Par.& Trt. Seq. 1& Trt. Seq. 2 & Trt. Seq. 3& Trt. Seq. 4&Trt. Seq. 5&Trt. Seq. 6\\
\toprule
$\beta_0$ & 0.00 (0.06) & 0.01 (0.12) & 0.01 (0.12) & 0 (0.06) & 0.02 (0.11) & 0.02 (0.11)\\

$\beta_1$ & 0.00 (0.09) & 0.02 (0.2) & 0.00 (0.35) & 0.00 (0.09) & -- (--) & -- (--)\\

$\beta_2$ & -- (--) & -- (--) & -- (--) & 0.04 (0.07) & 0.02 (0.17) & 0.06 (0.28)\\

$\beta_3$ & -- (--) & 0.01 (0.29) & 0.01 (0.29) & -- (--) & 0.06 (0.29) & 0.06 (0.29)\\
$\beta_{13}$&-- (--) & 0.02 (0.45) & 0.02 (0.45) & -- (--) & -- (--) & -- (--)\\
$\beta_{23}$&-- (--) & -- (--) & -- (--) & -- (--) & 0.07 (0.42) & 0.07 (0.42)\\
\bottomrule
\end{tabular*}
\centering
\begin{tabular*}{420pt}{@{\extracolsep\fill}cccccccc@{\fill}}
&\multicolumn{6}{@{}c@{}}{Correlation = 0.5}\\
\toprule
  Par.& Trt. Seq. 1& Trt. Seq. 2 & Trt. Seq. 3& Trt. Seq. 4&Trt. Seq. 5&Trt. Seq. 6\\
  \toprule
$\beta_0$ & 0.00 (0.06) & 0.01 (0.11) & 0.01 (0.11) & 0.00 (0.06) & 0.01 (0.11) & 0.01 (0.11)\\
$\beta_1$ & 0.01 (0.09) & 0.02 (0.19) & 0.04 (0.39) & 0.01 (0.09) & -- (--) & -- (--)\\
$\beta_2$ & -- (--) & -- (--) & -- (--) & 0.07 (0.09) & 0.01 (0.17) & 0.08 (0.28)\\
$\beta_3$ & -- (--) & 0.01 (0.28) & 0.01 (0.28) & -- (--) & 0.06 (0.32) & 0.06 (0.32)\\
$\beta_{13}$ & -- (--) & 0.02 (0.43) & 0.02 (0.43) & -- (--) & -- (--) & -- (--)\\
$\beta_{23}$& -- (--) & -- (--) & -- (--) & -- (--) & 0.06 (0.44) & 0.06 (0.44)\\
\bottomrule
\end{tabular*}
\centering
\begin{tabular*}{420pt}{@{\extracolsep\fill}cccccccc@{\fill}}
&\multicolumn{6}{@{}c@{}}{Correlation = 0.8}\\
\toprule
  Par.& Trt. Seq. 1& Trt. Seq. 2 & Trt. Seq. 3& Trt. Seq. 4&Trt. Seq. 5&Trt. Seq. 6\\
  \toprule
$\beta_0$ & 0.00 (0.06) & 0.01 (0.10) & 0.01 (0.10) & 0.00 (0.06) & 0.01 (0.10) & 0.01 (0.10)\\
$\beta_1$ & 0.01 (0.09) & 0.02 (0.18) & 0.02 (0.41) & 0.01 (0.09) & -- (--) & -- (--)\\
$\beta_2$ & -- (--) & -- (--) & -- (--) & 0.09 (0.07) & 0.00 (0.16) & 0.09 (0.29)\\
$\beta_3$ & -- (--) & 0.01 (0.30) & 0.01 (0.30) & -- (--) & 0.07 (0.38) & 0.07 (0.38)\\
$\beta_{13}$&-- (--) & 0.01 (0.40) & 0.01 (0.40) & -- (--) & -- (--) & -- (--)\\
$\beta_{23}$&-- (--) & -- (--) & -- (--) & -- (--) & 0.06 (0.46) & 0.06 (0.46)\\
\bottomrule
\end{tabular*}
\caption{ ENGAGE interaction model simulation results for $n=250$ sample size and approximately $200$ replicates. $\beta_0$ is the intercept; $\beta_1$ is the coefficient corresponding to $D_{11}$ (stage-1 MI-IOP); $\beta_2$ is the coefficient corresponding to $D_{12}$ (stage-1 MI-PC); $\beta_3$ is the coefficient corresponding to $D_{22}$; $\beta_{13}$ is the coefficient corresponding to the interaction between $D_{11}$ and $D_{22}$; and $\beta_{23}$ is the coefficient corresponding to the interaction between $D_{12}$ and $D_{22}$ (stage-2 MI-PC). -- = not part of model.}
\label{tab:ENGAGESimResultsInteraction}
\end{table*}

We compare the main effects model fit with the interaction model fit for the following treatment sequence outcomes: \begin{align*}
    Y_{i1} &=0.7+0.6D_{11i}+\epsilon_{1i}&Y_{i4} &=0.7+0.6D_{11i}+0.6D_{12i}+\epsilon_{4i}\\
    Y_{i2}&=0.2+0.7D_{11i}+0.9D_{22i}+2.0D_{11i}D_{22i}+\epsilon_{2i}&Y_{i5} &=0.3+0.6 D_{12i}+0.7D_{22i}+1.5D_{12i}D_{22i}+\epsilon_{5i}\\
    Y_{i3}&=0.2+0.6D_{11i}+0.9D_{22i}+2.0D_{11i}D_{22i}+\epsilon_{3i}&Y_{i6} &=0.3+0.6D_{12i}+0.7D_{22i}+1.5D_{12i}D_{22i}+\epsilon_{6i}
\end{align*} where $\epsilon_{ki}\sim \mathcal{N}(0,\sigma^2_k), k=1,...,6$. We impose the additional equality constraints that $\beta_4^{(3)}=\beta_4^{(2)}$ and $\beta_4^{(6)}=\beta_4^{(5)}$ for the interaction model. The results are summarized in Table \ref{tab:ENGAGESimResultsInteraction}. The standard errors are inflated due to collinearity induced by including the interaction terms. However, the results have low bias.
Note that the Watanabe–Akaike information criterion is lower for the interaction fit when the true model is the interaction model (see Table \ref{tab:WAICSimENGAGE}). However, for treatment sequence 3, the model's Watanabe–Akaike information criterion is only slightly lower for the interaction model.

\subsection{General SMART Study}
\begin{table*}[t]
\centering
\small
        \begin{center}
\begin{tabular*}{470pt}{cccccccccc@{\fill}}
\toprule
&\multicolumn{8}{@{}c@{}}{Bias (Standard Errors) for Main Effects Model}\\
\toprule
&\multicolumn{8}{@{}c@{}}{Correlation = 0.2}\\
\toprule
  Par.& Seq. 1& Seq. 2 & Seq. 3& Seq. 4&Seq. 5&Seq. 6&Seq. 7&Seq. 8\\
\toprule
$\beta_0$ & 0.01 (0.07) & 0.01 (0.09) & 0.00 (0.06) & 0.00 (0.06) & 0.02 (0.09) & 0.00 (0.12) & 0.00 (0.05) & 0.00 (0.05)\\

$\beta_1$ & 0.01 (0.10) & 0.01 (0.11) & 0.00 (0.09) & 0.00 (0.09) & -- (--) & -- (--) & -- (--) & -- (--)\\

$\beta_2$ & -- (--) & 0.00 (0.09) & -- (--) & -- (--) & -- (--) & 0.00 (0.12) & 0.00 (0.08) & -- (--)\\

$\beta_3$ & -- (--) & -- (--) & 0.01 (0.09) & 0.01 (0.09) & -- (--) & -- (--) & -- (--) & 0.00 (0.08)\\
$\beta_{4}$&-- (--) & -- (--) & -- (--) & 0.01 (0.14) & -- (--) & -- (--) & -- (--) & 0.01 (0.14)\\
$\beta_5$&0.00 (0.07) & -- (--) & -- (--) & -- (--) & 0.02 (0.14) & 0.00 (0.15) & 0.00 (0.08) & 0.00 (0.08)\\
\bottomrule
\end{tabular*}
\caption{ General SMART simulation results for main effects only model with $n=250$ sample size and approximately $200$ replicates for General. -- = not part of model.}
\label{tab:GeneralSimResultsMainEffects}
\end{center}

\end{table*}
 Next, we summarize the general SMART (see Figure \ref{fig:General-SMART} for the study design). This SMART has the most general structure of SMART designs we consider in this article. Individuals are initially randomized to Treatments (Trt.) 1 and 2. After the a set period of time (e.g., 2 weeks), subjects are subsequently monitored weekly for non-response to determine stage-2 treatments. 
 
 If a subject is classified as a non-responder, they move to stage 2 and are re randomized to one of two treatments: switch from Trt. 1 (or 2) to Trt. 4 or add Trt. 4 to Trt. 1 (or 2). At week 8, for example, individuals who do not meet their assigned non-response criterion are considered responders and re-randomized to one of two interventions: add Trt. 3 to Trt. 1 (or 2) or continue Trt. 1 (or 2) alone.

We model 5 potential compliances in the General SMART corresponding to the following interventions: Trt. 1, Trt. 2, Trt. 1+Trt 3 or Trt 2 + Trt 3, Trt 1+Trt 4 or Trt 2+Trt 4, Trt 4. There are a total of 8 treatment sequences and 8 embedded DTRs. For example, one of the embedded DTRs is given by the following decision rules: start the treatment with Trt. 1. If an individual becomes a non-responder, add Trt. 4; if at week 8 the individual is classified as a
responder, add Trt. 3. See Table \ref{tab:generalsmartEDTR} for the other embedded DTRs.

\subsection{General SMART simulation}

 The potential compliances were generated from a Gaussian copula model using the \texttt{R} package \texttt{copula}. This model is fully specified by the margins and correlation matrix. We use Beta distributed margins such that the parameters for the stage-1 compliances are $(3,2)$ and the stage-2 compliances are $(2,1)$, $(2,3)$, $(2,1)$, $(3,2)$. The correlation between potential compliances was set to $0.2$. 
We generated stage-1 response indicators as $S_i \sim \mathrm{Bern}\left\{\dfrac{\exp(D_{11}-1)}{1+\exp(D_{11}-1)}\right\}$  when  $A_{1i}=+1$ and  $S_i \sim \mathrm{Bern}\left\{\dfrac{\exp(D_{12}-1.5)}{1+\exp(D_{12}-1.5)}\right\}$ when $A_{1i}=-1$.
The following potential outcome models were generated for each of the 6 treatment sequences embedded in the SMART. 
\begin{align*}
    Y_{i1} &=1.0+0.6D_{11i}+\epsilon_{1i}&Y_{i5} &=0.7+0.6D_{12i}+\epsilon_{5i}\\
    Y_{i2} &= 0.4+0.5D_{11i}+0.8D_{22i}^{\mathrm{R}}+\epsilon_{2i}&Y_{i6} &= 0.6+0.2D_{12i}+0.4D_{22i}^{\mathrm{R}}+\epsilon_{6i}\\
    Y_{i3} &= 0.2 + 0.8D_{11i}+0.9D_{21i}^{\mathrm{NR}}+\epsilon_{3i}&Y_{i7} &= 0.4 + 0.5D_{12i}+0.9D_{21i}^{\mathrm{NR}}+\epsilon_{7i}\\
    Y_{i4} &= 0.2+0.8D_{11i}+0.9D_{21i}^{\mathrm{NR}}+0.7D_{22i}^{\mathrm{NR}}+\epsilon_{4i}&Y_{i8} &= 0.4+0.5D_{12i}+0.9D_{21i}^{\mathrm{NR}}+0.7D_{22i}^{\mathrm{NR}}+\epsilon_{8i}    
\end{align*}
where $\epsilon_{ki}\overset{iid}{\sim} \mathcal{N}(0,0.1^2)$.
The results are summarized in Table \ref{tab:GeneralSimResultsMainEffects}. Similarly to the ENGAGE simulation, the results for EXTEND indicate low bias and standard errors. The low standard errors is despite there being fewer subjects in each of the treatment sequences when compared with ENGAGE.
\section{Real data application: ENGAGE study}

We have assessed the performance of our method on simulated versions of the ENGAGE study. Now, we apply our methodology to the real data in order to determine which EDTRs are optimal across compliance classes. Since the data is longitudinal, we define our outcome as the log of the sum of days from weeks 2 to 24 in which alcohol was consumed and the sum of the days cocaine was consumed plus a small positive constant. We consider only African-Americans who did not engage by week 2. We fit the same models as in the simulation studies. We do not incorporate covariates besides potential compliances in our analysis.
\afterpage{\begin{table}
\centering
\begin{tabular}[t]{ccccccc}
\toprule
&\multicolumn{5}{@{}c@{}}{Watanabe–Akaike information criterion for Simulated ENGAGE}\\
\toprule
Model &Trt. Seq. 1&Trt. Seq. 2& Trt. Seq. 3& Trt. Seq. 4& Trt. Seq. 5 &Trt. Seq. 6\\
\toprule
Main Effects & -80.63 & -44.28 & 32.32 & -38.88 & 516.80 & 555.14\\
\midrule
Interaction & -80.51 & -61.57 & 30.91 & -38.38 & 9.45 & 358.10\\
\bottomrule
\end{tabular}
\caption{Watanabe–Akaike information criterion also known as the widely applicable information criterion  for each treatment sequence in the simulated ENGAGE SMART for the main effects only model and the interaction model. The results are averaged across approximately $200$ simulation replicates.}
\label{tab:WAICSimENGAGE}
\end{table}
}

\begin{figure}[t]
\centering
\includegraphics[width=5in, trim = {0cm 0.2cm 0cm 1cm},clip=true]{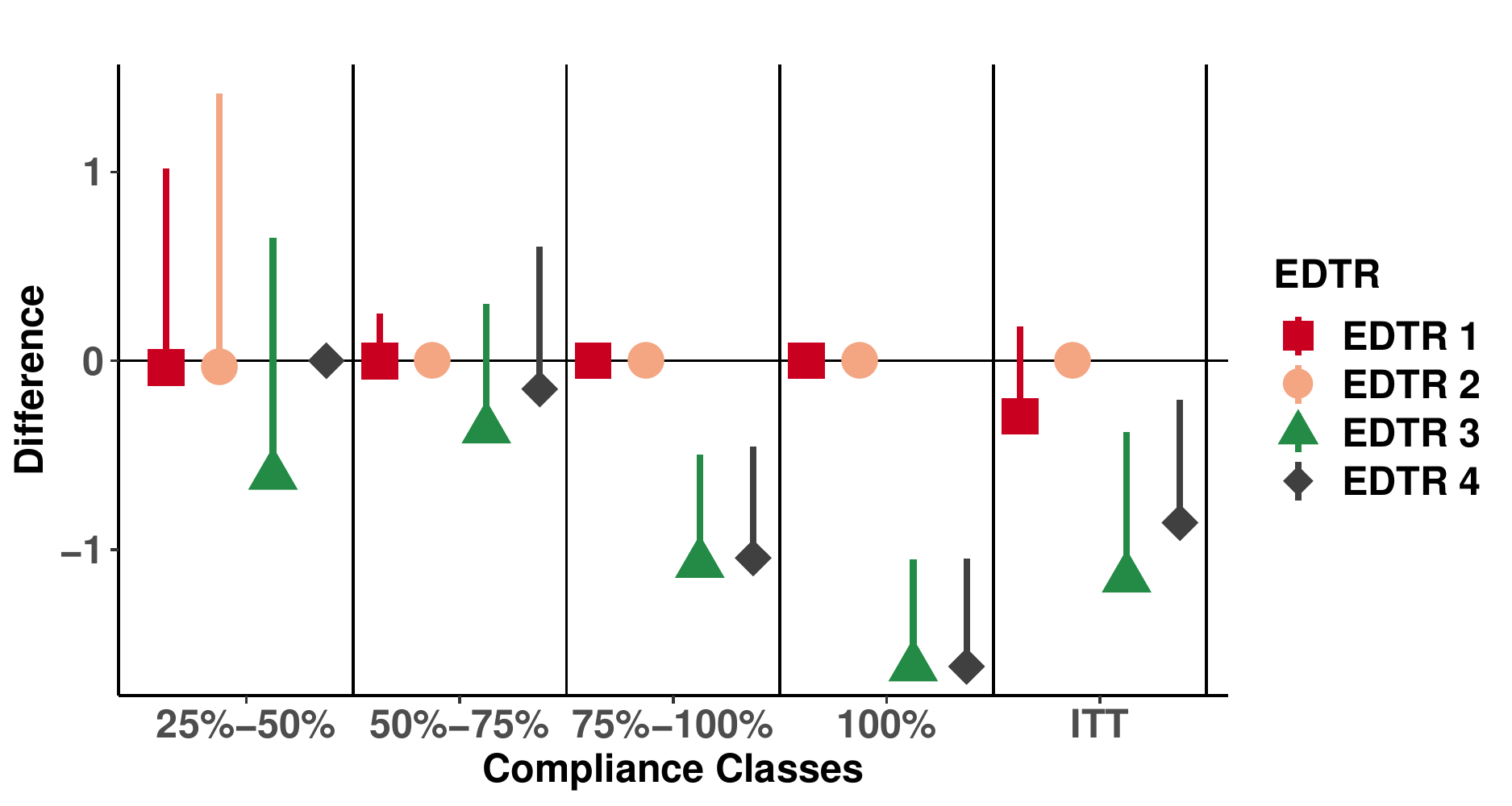}
\caption{MCB applied to the real ENGAGE SMART. See Table \ref{tab:EDTR-ENGAGE-table} for a legend of embedded DTRs. EDTR = embedded DTR, ITT = intention-to-treat. Simultaneous upper one-sided credible intervals by compliance class for the difference between each EDTR and the optimal embedded DTR. ITT are confidence intervals computed using the bootstrap.}
\label{fig:MCB-ENGAGE}
\end{figure}

In Figure \ref{fig:MCB-ENGAGE}, we apply MCB. Within each compliance class the median compliance in the quantile interval is chosen for each potential compliance. Intention-to-treat analysis was carried out using a marginal structural model \citep{nahum2012experimental}. Confidence intervals were constructed for the MSM using the bootstrap. We see that for all the compliance classes, embedded DTRs 1 and 2 are optimal. For compliance classes 75\%-100\%, 100\%, and intention-to-treat (ITT), 3 and 4 are significantly inferior to the optimal embedded DTR. In 25\%-50\% and 50\%-75\%, all embedded DTRs are optimal. Although not statistically significant, the embedded DTR in which patients receive NFC is more efficacious than subjects who receive MI-PC in intention-to treat.

\begin{figure}[t]
\centering
\includegraphics[width = 2.5in, trim = {0cm 2cm 0cm 0cm},clip=true]{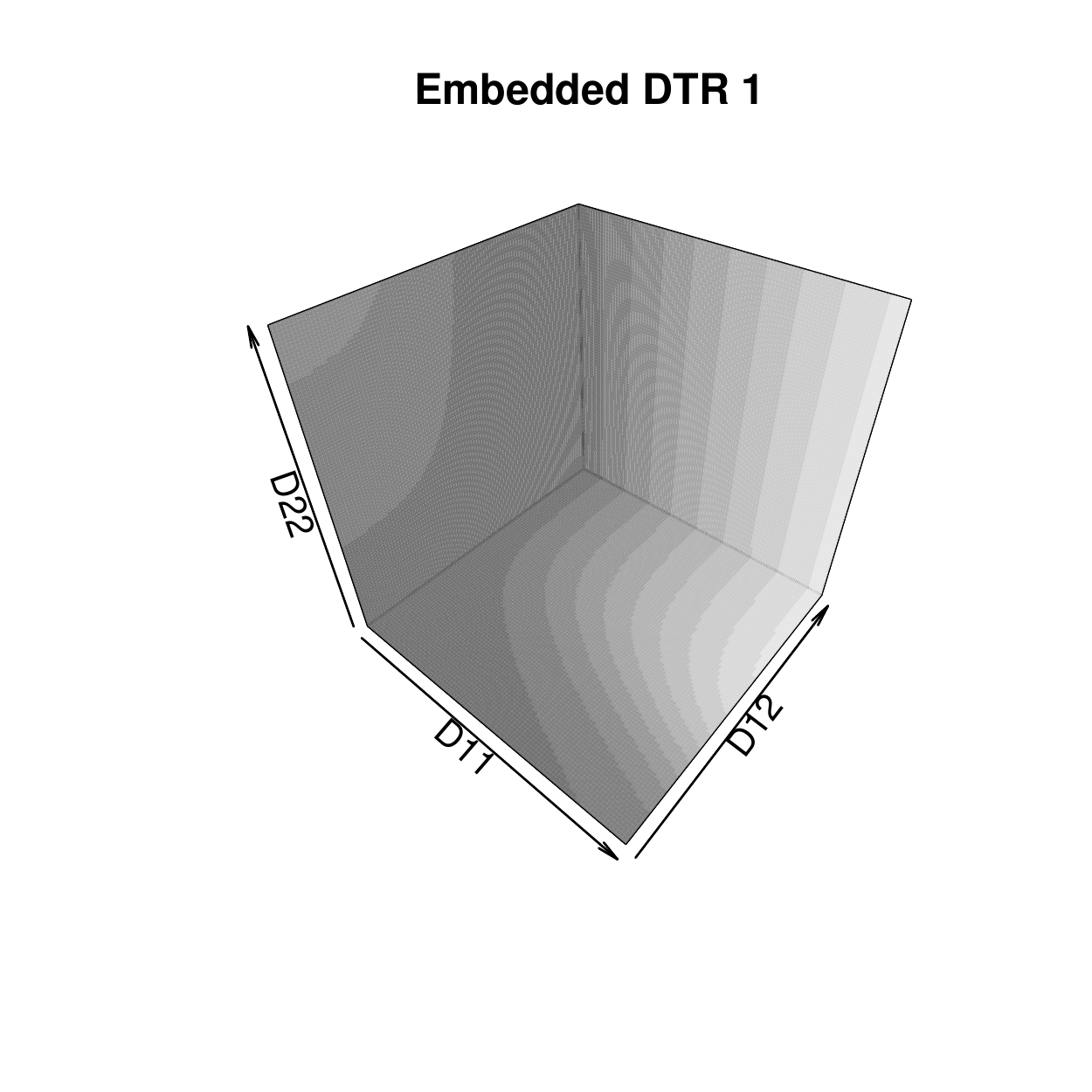}
\includegraphics[width = 2.5in, trim = {0cm 2cm 0cm 0cm},clip=true]{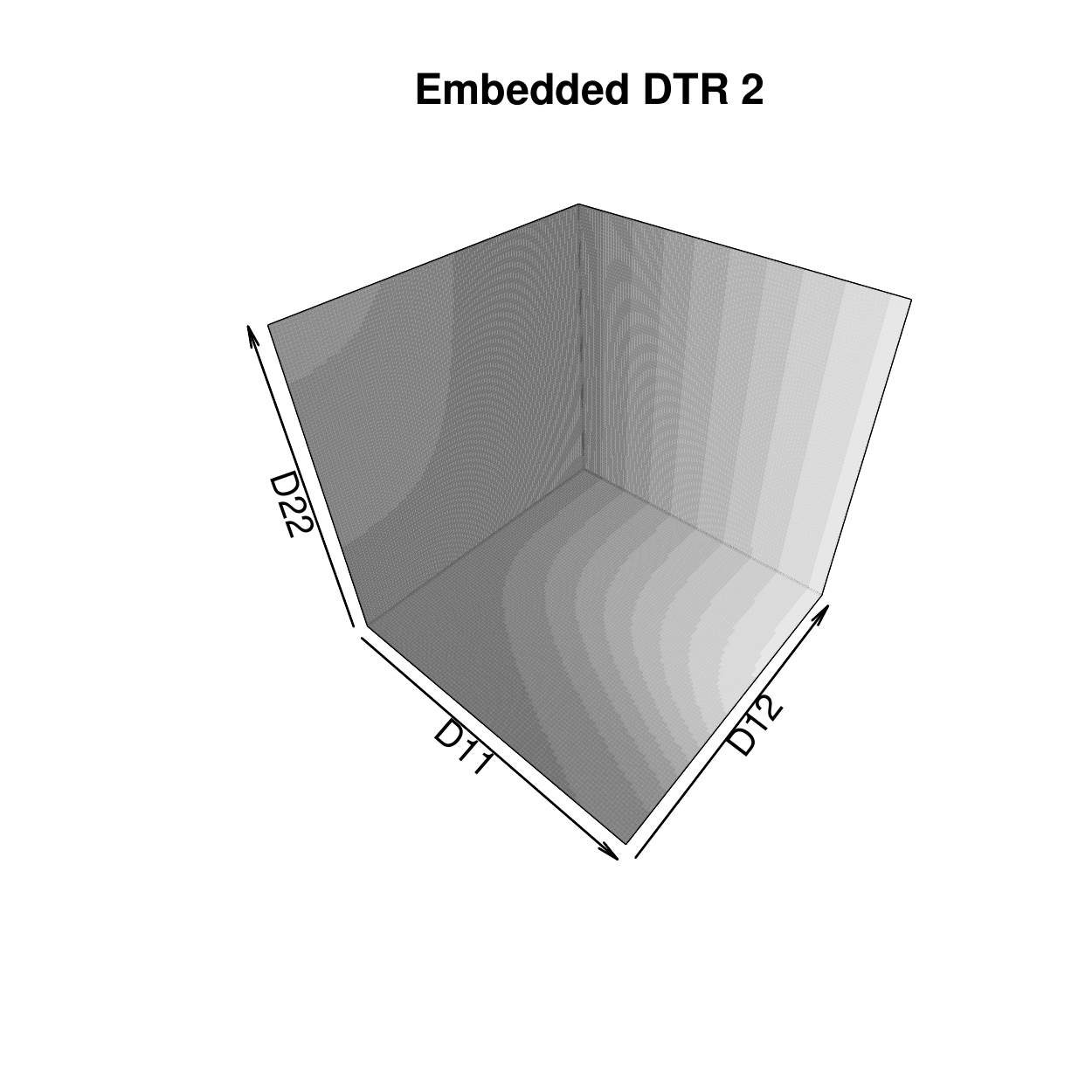}
\includegraphics[width = 2.5in, trim = {0cm 2cm 0cm 0cm},clip=true]{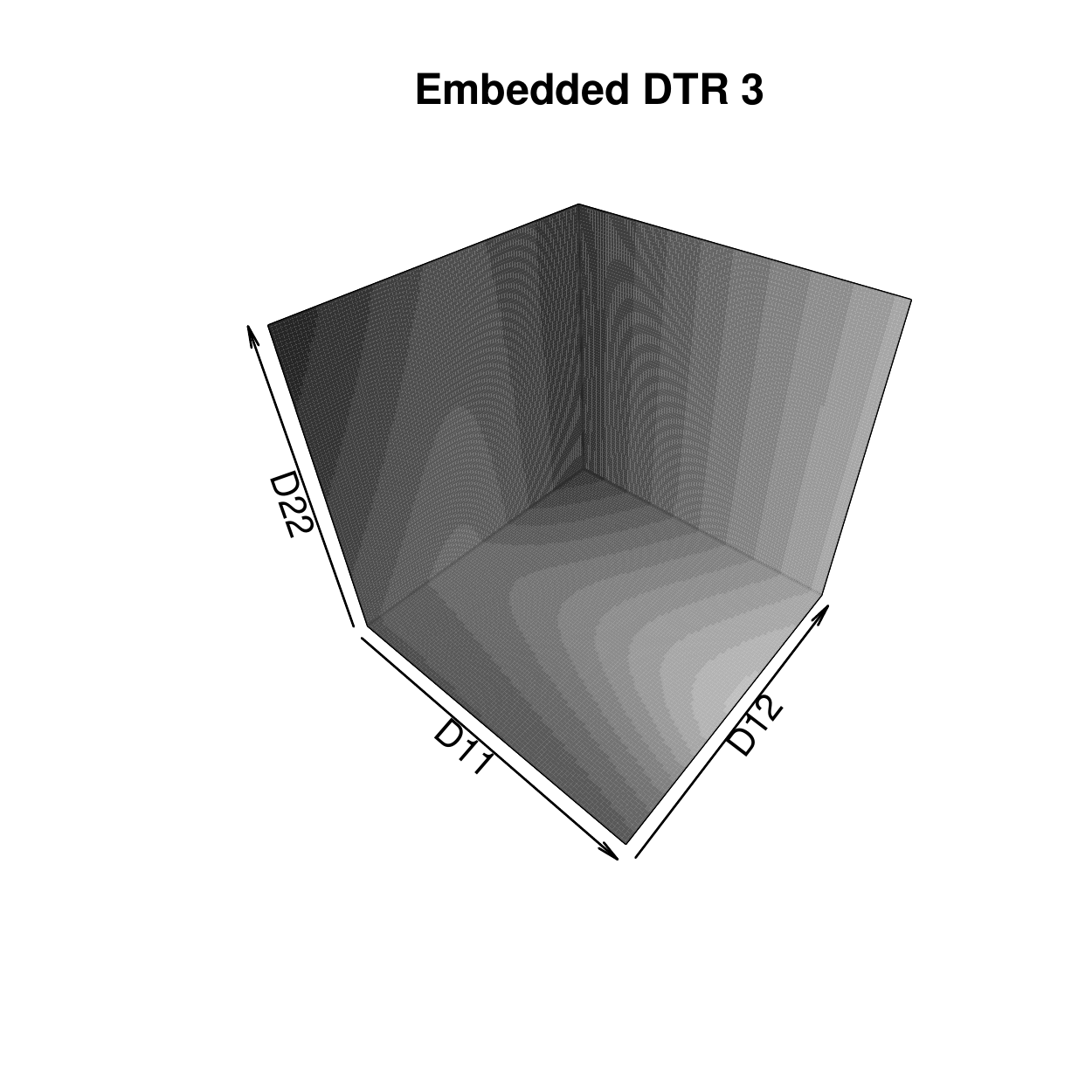}
\includegraphics[width = 2.5in, trim = {0cm 2cm 0cm 0cm},clip=true]{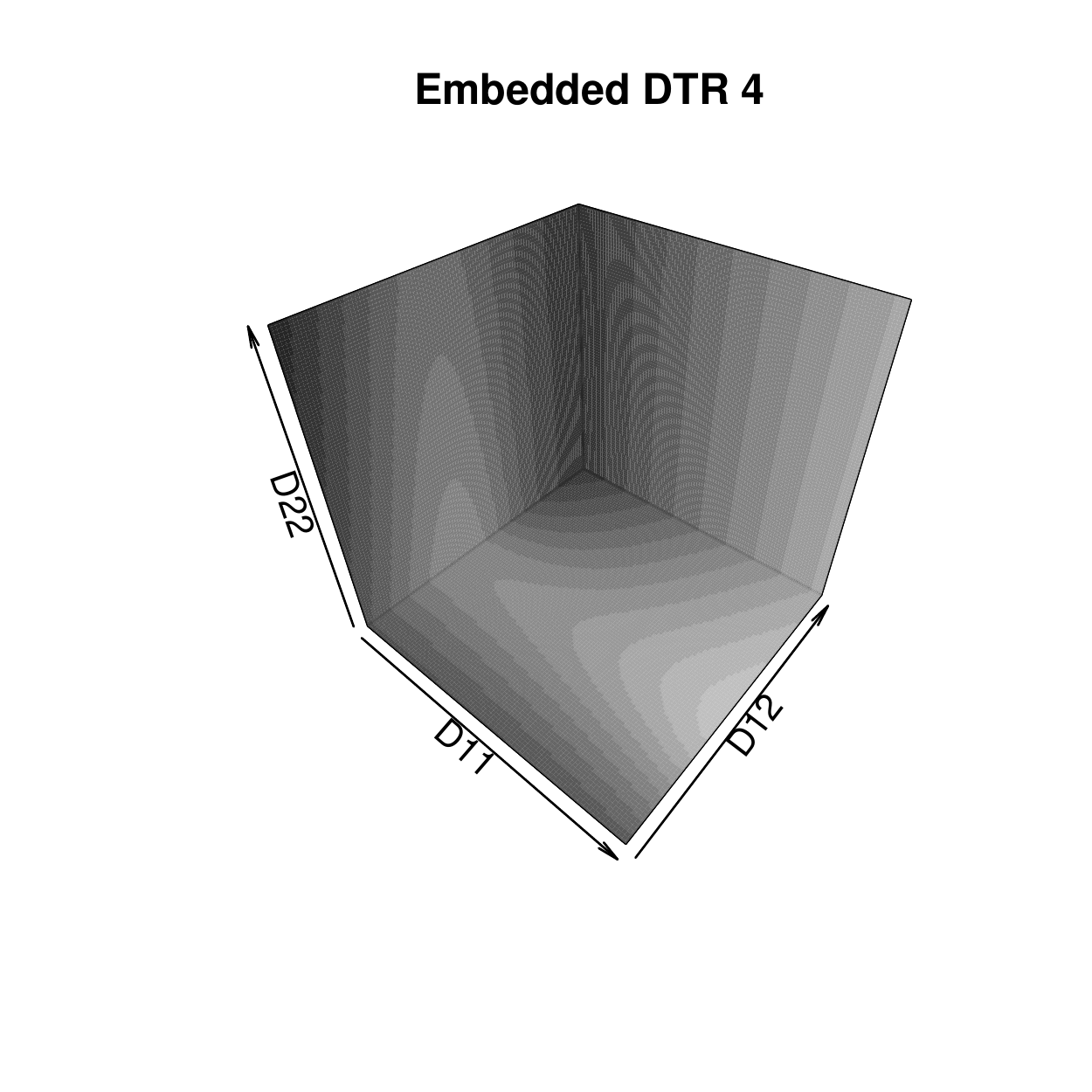}
\begin{center}\includegraphics[width = 3in, trim = {0cm 1cm 0cm 2cm},clip=true]{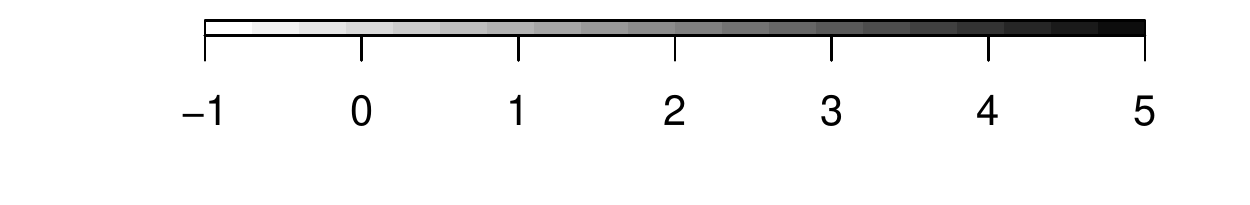}\end{center}
\caption{Real ENGAGE SMART. 3D representation of 4D surface by potential compliance plots. The shade represents the log of the sum of the number of days drinking alcohol and the number of days consuming cocaine. Lighter shades are more favorable.}
\label{fig:3d-plots}
\end{figure}

In Figure \ref{fig:3d-plots}, potential compliances in ENGAGE, $D_{11},D_{12},D_{22}$ are plotted with outcome indicated by shade. Lighter shades indicate more favorable outcomes. Note that as the potential compliances vary between $0$ and $1$, the probability of stage-1 response changes which affects how much weight each treatment sequence outcome contributes to the embedded DTR outcomes. This may make it so that the outcome is not strictly monotone in a potential compliance. Note that these plots suggest optimal embedded DTRs consist of MI-IOP in stage-1.

In Table \ref{tab:ENGAGE-Real-Main-Effects-Interaction}, we have point estimates and corresponding standard errors listed for each of the potential compliance regression coefficients. In Table \ref{tab:Real-Engage-MainEffects-EDTR-outcomes}, we have the individual treatment sequence point estimates and standard errors. We first summarize the main effects model. We see that for all treatment sequences higher first stage compliance leads to less days consuming alcohol/cocaine. For treatment sequences 2 and 3, there is some evidence that stage-2 potential compliance leads to better outcomes. For treatment sequences 5 and 6, higher potential compliances in stage 2 are associated with poor outcomes, but the standard errors for the corresponding coefficients are large. The interpretation of the interaction model may proceed similarly. However, despite the large negative interaction terms, the large standard error complicates the interpretation. In Table \ref{tab:WAIC-ENGAGE-Real}, we see that the Watanabe-Akaike information criterion suggests the Main Effects model is a better fit than the interaction model as it has lower information criterion values.

\section{Discussion}

Compliance is an important predictor of outcomes for substance use disorders commonly studied in SMARTs. In this paper, we fill the methodological gap in accounting for partial compliance in the SMART setting. Adjusting for partial as opposed to binary compliance allows for more flexibility when there is continuous compliance and the threshold for compliers vs. non-compliers is not easily determined. Challenges in adjusting for compliance include the post-treatment adjustment bias for observed compliances, and assumptions which must be imposed for identifiability. We presented sufficient conditions for identifiability of the parametric regression model.
We extend the methodology in \cite{schwartz2011bayesian} from single-stage decision to multi-stage decision settings, so that the embedded DTRs outcomes may be determined as a function of partial compliance.
We applied a semiparametric Bayesian model for principal stratification to determine the optimal embedded DTR for a given compliance class.

When choosing an optimal embedded DTR from the set of best constructed using MCB, one can use information about the compliance of a patient. If such information is not known, choosing a DTR which is in the set of best across a plausible range of compliance classes is reasonable. An alternative approach is to use demographic information to predict how compliant a patient will likely be; hence, which DTR is optimal. For example, one could consider whether a patient's age falls in a specific window.

The method leverages a non-parametric Dirichlet process mixture for kernel density estimation of the joint potential compliance distribution allowing for complex but realistic compliance distributions to be modeled. It does not require arbitrary dichotomization as it models partial compliance as a continuous variable between 0 and 1 which is particularly useful and realistic for conditions studied in SMARTs. Principal stratification elucidates the effect of compliance on outcomes and avoids the problem of adjusting for observed compliances which are post-treatment variables. Furthermore, we showed how MCB may be implemented in the Bayesian setting and how to apply it to construct sets of best embedded DTRs within compliance classes.
\afterpage{\begin{table*}[t]
\centering
\small
        \begin{center}
\begin{tabular*}{440pt}{cccccc}
\toprule
\multicolumn{6}{@{}c@{}}{Mean Embedded DTR Outcomes (Standard Errors) for Main Effects Model (Simulated)}\\
\toprule
Compliance Class & EDTR 1 & EDTR 2 & EDTR 3 & EDTR 4&Set of Best ($\hat{\mathcal{B}}$)\\
\toprule
25\%-50\% & 1.10 (0.01)& 1.17 (0.04)& 1.21 (0.01)& 1.26 (0.03) & EDTR 1, EDTR 2\\
\midrule
50\%-75\% & 1.45 (0.04)& 1.53 (0.05)& 1.59 (0.01)& 1.65 (0.02)& EDTR 1\\
\midrule
75\%-100\% & 1.79 (0.11) & 1.87 (0.13)& 2.00 (0.02)& 2.06 (0.04)& EDTR 1, EDTR 3\\
\midrule 
100\% & 2.15 (0.26)& 2.21 (0.29)& 2.53 (0.08)& 2.59 (0.09) &EDTR 1, EDTR 3, EDTR 4\\
\midrule
ITT & 1.19 (0.05)& 1.20 (0.05)& 1.37 (0.04)& 1.38 (0.05)& EDTR 1, EDTR 2\\
\bottomrule
\end{tabular*}
\caption{ Simulated ENGAGE SMART main effects only model EDTR point estimates by compliance class.}
\end{center}
        \label{tab:simulatedENGAGE-EDTR-OUTCOMES}
\end{table*}}

Previous work in two-arm clinical trials have failed to validate their method either analytically or through simulation. We demonstrated the validity of our method both mathematically and through extensive simulation studies for varying degrees of correlation between the potential compliances. While we focused on a two-stage SMART with at most two different interventions at each stage of randomization, the method may be extended to SMARTs with a greater number of potential compliances.

One of the major challenges with principal stratification, is that some of the potential compliances are latent. This is handled via data augmentation. By making identifiability assumptions similar to \cite{schwartz2011bayesian}, we were able to achieve low bias and small standard errors. Future work should entail tailoring treatments according to compliance using Q-learning and relaxing the linearity assumption in the outcome-compliance model. An alternative approach which still uses principal stratification would be to model each embedded DTR directly rather than each treatment sequence. The pitfall for such an approach is it would not allow as much flexibility when defining the identifiability constraints for the regression coefficients. If we could observe the potential compliances, then modeling the embedded DTRs directly would be straightforward using a weighted and replicated linear regression. However, some potential compliances are missing. Since the pattern of missingness is at the level of treatment sequence rather than embedded DTR, imputation of missing compliances is made more difficult with direct modeling.

Another important extension to the proposed method would be to account for partial or binary compliance in a way that fully accounts for the often longitudinal nature of compliance.

\clearpage
\bibliographystyle{authordate1}
\bibliography{main}
\clearpage
\afterpage{ \begin{table*}[t]
    \begin{adjustwidth}{-.5in}{-.5in} 

\setlength{\tabcolsep}{2pt}

\footnotesize
        \begin{center}
        
\begin{tabular*}{490pt}{cccccccccc}

\toprule
\multicolumn{10}{@{}c@{}}{Mean Embedded DTR Outcomes (Standard Errors) for Main Effects Model: General SMART}\\
\toprule
\makecell{Compliance \\Class} & EDTR 1 & EDTR 2 & EDTR 3 & EDTR 4 & EDTR 5 & EDTR 6 & EDTR 7 & EDTR 8&Set of Best ($\hat{\mathcal{B}}$)\\
\toprule
25\%-50\% & 1.05 (0.03)& 1.27 (0.02)& 0.99 (0.02) & 1.22 (0.02) & 0.96 (0.01) & 1.31 (0.02)& 0.94 (0.01)& 1.29 (0.02)& EDTRs 3, 5, 7\\
\midrule
50\%-75\% & 1.26 (0.02)& 1.56 (0.02)& 1.24 (0.02) & 1.54 (0.03)& 1.15 (0.01)& 1.58 (0.03) & 1.14 (0.02) & 1.57 (0.04)& EDTRs 5, 7\\
\midrule
75\%-100\% & 1.52 (0.02)& 1.90 (0.05) & 1.52 (0.02)& 1.90 (0.05)& 1.37 (0.02)& 1.87 (0.07) & 1.37 (0.02) & 1.87 (0.07)& EDTRs 5,7\\
\midrule 
100\% &1.87 (0.04)& 2.30 (0.10)& 1.86 (0.04) & 2.29 (0.11)& 1.62 (0.05)& 2.14 (0.12)& 1.63 (0.05) & 2.15 (0.12)&EDTRs 5, 7\\
\midrule
ITT & 1.31 (0.04) & 1.27 (0.04)& 1.29 (0.04)& 1.25 (0.04) & 1.18 (0.04) & 1.14 (0.04)& 1.16 (0.03)& 1.12 (0.03) & EDTRs 5, 6 ,7,8\\
\bottomrule
\end{tabular*}
\caption{ Simulated general SMART main effects only model EDTR point estimates by compliance class.}
\label{tab:SimGeneralSMARTMainEffectsEDTRSummary}
\end{center}
        \end{adjustwidth}

\end{table*}}
\afterpage{\begin{table*}[t]
\centering
\small
        \begin{center}
\begin{tabular*}{430pt}{cccccc}
\toprule
\multicolumn{6}{@{}c@{}}{Mean Embedded DTR Outcomes (Standard Errors) for Main Effects Model}\\
\toprule
Compliance Class & EDTR 1 & EDTR 2 & EDTR 3 & EDTR 4&Set of Best ($\hat{\mathcal{B}}$)\\
\toprule
25\%-50\% & 2.52 (0.14) & 2.51 (0.48)& 3.09 (0.36) & 2.48 (0.48) & EDTRs 1, 2, 3, and 4\\
\midrule
50\%-75\% & 1.95 (0.29) & 1.95 (0.31) & 2.31 (0.08) & 2.10 (0.12)& EDTRs 1, 2, 3, and 4\\
\midrule
75\%-100\% & 0.50 (0.28) & 0.50 (0.28)& 1.57 (0.05) & 1.54 (0.05)&EDTR 1, EDTR 2\\
\midrule 
100\% & -0.18 (0.26)& -0.18 (0.27)& 1.45 (0.05)& 1.44 (0.05)&EDTR 1, EDTR 2\\
\midrule
ITT & 2.02 (0.26)& 1.73 (0.25) & 2.88 (0.22) & 2.59 (0.22)&EDTR 1, EDTR 2\\
\bottomrule
\end{tabular*}
\caption{ Real ENGAGE SMART main effects only model EDTR point estimates by compliance class.}
        \label{tab:Real-Engage-MainEffects-EDTR-outcomes}

\end{center}
\end{table*}}

\begin{table*}[t]
\centering

        \begin{center}
\begin{tabular*}{410pt}{ccccccccc@{\fill}}
\toprule
&\multicolumn{6}{@{}c@{}}{Point Estimates (Standard Errors) for Main Effects Model}\\
\toprule
  Par.& Seq. 1& Seq. 2 & Seq. 3& Seq. 4&Seq. 5&Seq. 6\\
\toprule
$\beta_0$ & 2.66 (0.44) & 2.77 (0.86) & 2.77 (0.86) & 2.66 (0.44) & 3.49 (0.46) & 3.49 (0.46)\\

$\beta_1$ & -2.76 (0.75) & 0.06 (1.78) & -2.43 (3.05) & -2.76 (0.75) & -- (--) & -- (--)\\

$\beta_2$ & -- (--) & -- (--) & -- (--) & -0.35 (1.16) & -1.61 (1.68) & -4.46 (2.67)\\
$\beta_3$ & -- (--) & -2.16 (2.11) & -2.16 (2.11) & -- (--) & 0.82 (0.98) & 0.82 (0.98)\\
\bottomrule
&\multicolumn{6}{@{}c@{}}{Point Estimates (Standard Errors) for Interaction Model}\\
\toprule
  Par.& Seq. 1& Seq. 2 & Seq. 3& Seq. 4&Seq. 5&Seq. 6\\
\toprule
$\beta_0$ & 2.67 (0.44) & 3.01 (1.00) & 3.01 (1.00) & 2.67 (0.44) & 3.57 (0.61) & 3.57 (0.61)\\

$\beta_1$ & -2.78 (0.75) & -0.64 (2.3) & -2.7 (4.28) & -2.78 (0.75) & -- (--) & -- (--)\\

$\beta_2$ & -- (--) & -- (--) & -- (--) & 0.37 (0.72) & -2.28 (3.16) & -5.74 (3.82)\\

$\beta_3$ & -- (--) & -4.23 (4.13) & -4.23 (4.13) & -- (--) & 0.54 (1.43) & 0.54 (1.43)\\
$\beta_{13}$ & -- (--) & 5.95 (9.82) & 5.95 (9.82) & -- (--) & -- (--) & -- (--)\\
$\beta_{23}$ & -- (--) & -- (--) & -- (--) & -- (--) & 1.33 (4.53) & 1.33 (4.53)\\
\bottomrule
\end{tabular*}
\caption{ Real data illustration: ENGAGE SMART. -- = not part of model.}
\label{tab:ENGAGE-Real-Main-Effects-Interaction}
\end{center}
        
\end{table*}

\afterpage{
\begin{table}
\centering
\begin{tabular}[t]{ccccccc}
\toprule
&\multicolumn{5}{@{}c@{}}{Real Data Illustration: Watanabe–Akaike information criterion}\\
\toprule
Model &Trt. Seq. 1&Trt. Seq. 2& Trt. Seq. 3& Trt. Seq. 4& Trt. Seq. 5 &Trt. Seq. 6\\
\toprule
Main Effects & 92.47 & 79.64 & 84.82 & 149.78 & 59.77 & 74.15\\
\midrule
Interaction & 92.59 & 82.71 & 86.35 & 148.30 & 63.66 & 74.45\\
\bottomrule
\end{tabular}
\caption{Watanabe–Akaike information criterion for each treatment sequence in the real ENGAGE SMART for the main effects only model and the interaction model.}
\label{tab:WAIC-ENGAGE-Real}
\end{table}

}

\clearpage
\clearpage

\section*{Appendix A: Representation of the PCE}

\begin{align*}
    \mathrm{PCE}^{(l)}(\bD)&=\E[Y^{(l)}\mid \boldsymbol{D} ]\\&=\E[Y \mid \EDTR_l,\boldsymbol{D}]\\  
    &= \E[Y \mid A_1=a_{1l},\EDTR_l,\boldsymbol{D}]\\
    &=\E[Y\mid A_1=a_{1l},S=1,\EDTR_l,\boldsymbol{D}]\Pr(S=1\mid A_1=a_{1l},\EDTR_l,\boldsymbol{D})\\&+\E[Y\mid A_1=a_{1l},S=0,\EDTR_l,\boldsymbol{D}]\Pr(S=0\mid A_1=a_{1l},\EDTR_l,\boldsymbol{D})\\
    &=\E[Y\mid A_{1}=a_{1l}, S=1,\boldsymbol{D}]\Pr(S=1\mid A_1=a_{1l},\boldsymbol{D})\\&+\E[Y\mid A_1=a_{1l},S=0,A_{2}^{\mathrm{NR}}=a_{2l},\boldsymbol{D}]\Pr(S=0\mid A_1=a_{1l},\boldsymbol{D})
\end{align*}
where $\EDTR_l$ is the $l^{th}$ embedded DTR.
\clearpage
\subsection*{Appendix B: Identification of Regression Coefficients}
Without loss of generality, we prove identifiability for a specific treatment sequences in the ENGAGE SMART for illustration.
Treatment sequences 1, 2, and 5 are fully observable so the regression coefficients are identified. 

Consider the model for treatment sequence $4$: $$Y_4=\beta_0^{(4)}+\beta_1^{(4)}D_{11}+\beta_2^{(4)}D_{12}.$$ Suppose assumptions A.2 (identifiability) and A.3 (homogeneity) hold. Then, $\beta_0^{(4)}=\beta_0^{(1)}$ and $\beta_1^{(4)}=\beta_1^{(1)}$. We will show that the coefficient corresponding to the observed compliance, $\beta_2^{(4)}$, is identified. Let $\bX_c$ denote the true design matrix and $\bX$ denote the design matrix that was imputed using data augmentation.

Write the true model in matrix notation as
$$\E[\boldsymbol{Y}\mid \bX_c]=\bX_c\bbeta^{(4)}$$

We have that 

\begin{align}
\hat{\bbeta}^{(4)}&=(\bX^{\top}\bX)^{-1}\bX^{\top}\boldsymbol{Y}\Rightarrow\\
\bX^{\top}\bX\hat{\bbeta}^{(4)}&=\bX^{\top}\boldsymbol{Y}\\&=\bX^{\top}\bX_c\bbeta^{(4)}\nonumber
\end{align}

Note that $$\boldsymbol{e}_1 \bX^{\top}\bX = \begin{pmatrix}\boldsymbol{1}^{\top}\boldsymbol{1}&\boldsymbol{1}^{\top}\tilde{\bD}_{11}&\boldsymbol{1}^{\top}\bD_{12}\end{pmatrix}$$

where $\tilde{\bD}_{11}$ denotes an imputed potential compliance.

and 

\begin{equation}\boldsymbol{e}_1 \bX^{\top}\bX_c = \begin{pmatrix}\boldsymbol{1}^{\top}\boldsymbol{1}&\boldsymbol{1}^{\top}{\bD}_{11}&\boldsymbol{1}^{\top}\bD_{12}\end{pmatrix}\end{equation}
where 
\begin{equation}
\boldsymbol{e}_1=\begin{pmatrix}1&0&0\end{pmatrix}
\end{equation}
Hence,

$$\boldsymbol{1}^{\top}\boldsymbol{1}\hat{\beta}_0^{(4)}+\boldsymbol{1}^{\top}\tilde{\bD}_{11}\hat{\beta}_1^{(4)}+\boldsymbol{1}^{\top}\bD_{12}\hat{\beta}_2^{(4)}\\=\boldsymbol{1}^{\top}\boldsymbol{1}\beta_0^{(4)}+\boldsymbol{1}^{\top}\bD_{11}\beta_1^{(4)}+\boldsymbol{1}^{\top}\bD_{12}\beta_2^{(4)}$$
by equations 6, 7, and 8.
Then, take the conditional expectation of both sides conditioning on $\boldsymbol{D}_{12}$, $\hat{\beta}^{(4)}_0=\beta_0^{(1)}$, $\hat{\beta}_1^{(4)}=\beta_1^{(1)}$, and the treatment sequence potential outcome $Y_4$.

$$\begin{aligned}
&\boldsymbol{1}^{\top}\boldsymbol{1}\E[\hat{\beta}_0^{(4)}\mid \bD_{12},\hat{\beta}_0^{(4)}=\beta_0^{(1)},\hat{\beta}_1^{(4)}=\beta_1^{(1)},Y_4]\\&+\boldsymbol{1}^{\top}\E[\tilde{\bD}_{11}\hat{\beta}_1^{(4)}\mid \bD_{12},\hat{\beta}_0^{(4)}=\beta_0^{(1)},\hat{\beta}_1^{(4)}=\beta_1^{(1)},Y_4]\\&+\boldsymbol{1}^{\top}\bD_{12}\E[\hat{\beta}_2^{(4)}\mid \bD_{12},\hat{\beta}_0^{(4)}=\beta_0^{(1)},\hat{\beta}_1^{(4)}=\beta_1^{(1)},Y_4]=\boldsymbol{1}^{\top}\boldsymbol{1}\beta_0^{(1)}+\boldsymbol{1}^{\top}\bD_{11}\beta_1^{(1)}+\boldsymbol{1}^{\top}\bD_{12}\beta_2^{(4)}
\end{aligned}$$

Hence, 
$$\begin{aligned}&\boldsymbol{1}^{\top}\boldsymbol{1}\beta_0^{(1)}\\&+\boldsymbol{1}^{\top}\beta_1^{(1)}\E[\tilde{\bD}_{11}\mid \bD_{12},\hat{\beta}_0^{(4)}=\beta_0^{(1)},\hat{\beta}_1^{(4)}=\beta_1^{(1)},Y_4]\\&+\boldsymbol{1}^{\top}\bD_{12}\E[\hat{\beta}_2^{(4)}\mid \bD_{12},\hat{\beta}_0^{(4)}=\beta_0^{(1)},\hat{\beta}_1^{(4)}=\beta_1^{(1)},Y_4]\\&=\boldsymbol{1}^{\top}\boldsymbol{1}\beta_0^{(1)}+\boldsymbol{1}^{\top}\bD_{11}\beta_1^{(1)}+\boldsymbol{1}^{\top}\bD_{12}\beta_2^{(4)}\end{aligned}$$

where we have used the identifiability constraints $\hat{\beta}_0^{(4)}=\beta_0^{(1)}$ and $\hat{\beta}_1^{(4)}=\beta_1^{(1)}$.

Rearranging yields:
$$\begin{aligned}&\E[\hat{\beta}_2^{(4)}-\beta_2^{(4)}\mid \bD_{12},\hat{\beta}_0^{(4)}=\beta_0^{(1)},\hat{\beta}_1^{(4)}=\beta_1^{(1)},Y_4]\\&=
\beta_1^{(1)}\dfrac{\boldsymbol{1}^{\top}}{\boldsymbol{1}^{\top}\bD_{12}}\left(\E[\bD_{11}\mid \bD_{12}]-\E[\tilde{\bD}_{11}\mid \bD_{12},\hat{\beta}_0^{(4)}=\beta_0^{(1)},\hat{\beta}_1^{(4)}=\beta_1^{(1)},Y_4]\right)\end{aligned}$$

Then, $$\E[\bD_{11}\mid \bD_{12}]-\E[\tilde{\bD}_{11}\mid \bD_{12},\hat{\beta}_0^{(4)}=\beta_0^{(1)},\hat{\beta}_1^{(4)}=\beta_1^{(1)},Y_4]=0$$
using well-known facts about the conditional distribution of multivariate normal distributions.

Hence, $\hat{\beta}_2^{(4)}$ is unbiased for $\beta_2^{(4)}$.
\clearpage
\subsection{Appendix C: Plots for simulation}
\begin{figure}[h]
\centering
\includegraphics[width = 5.8in, trim = {0cm 0cm 0cm 0cm},clip=true]{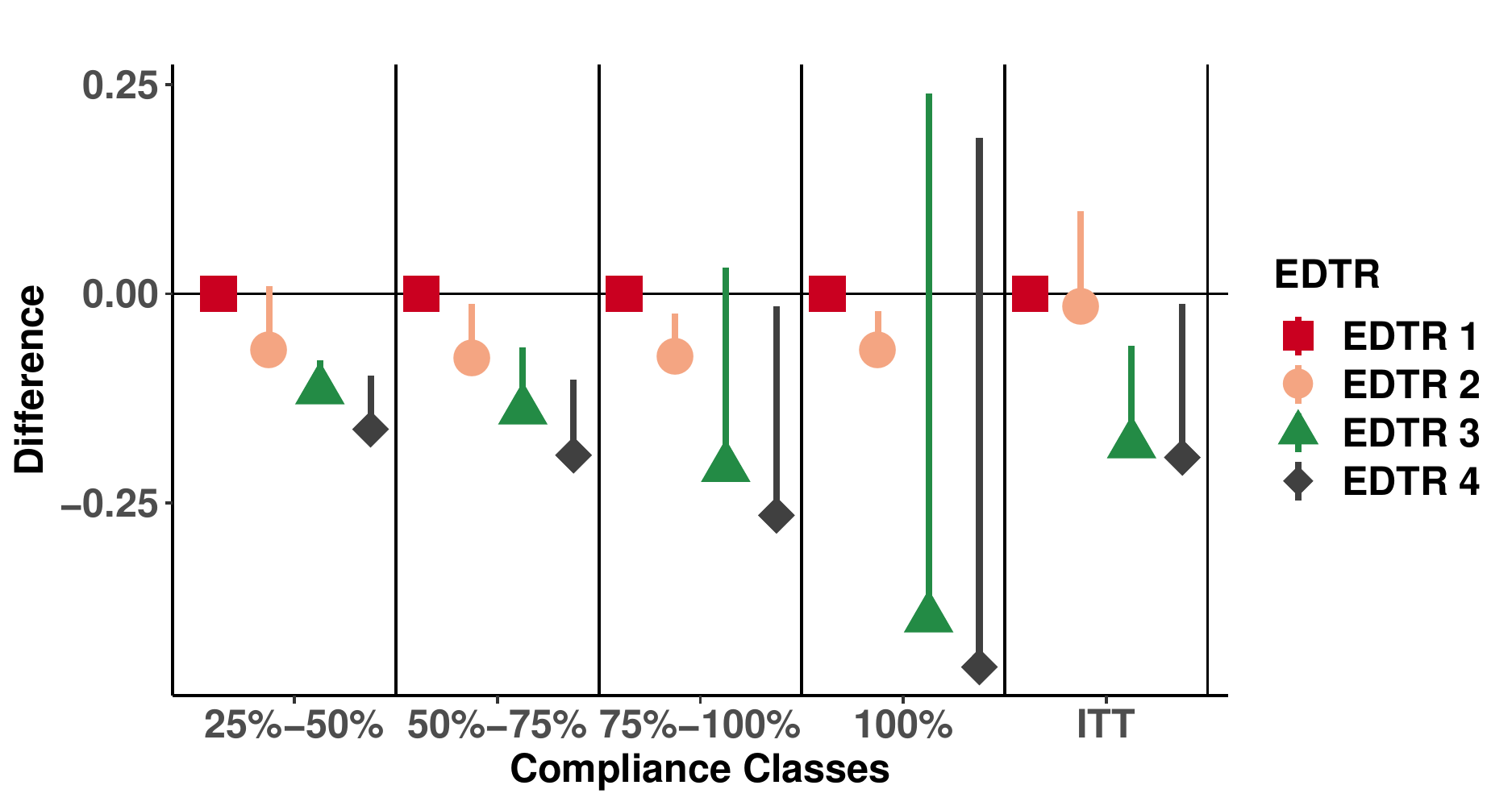}
\includegraphics[width = 5.8in, trim = {0cm 0cm 0cm 0cm},clip=true]{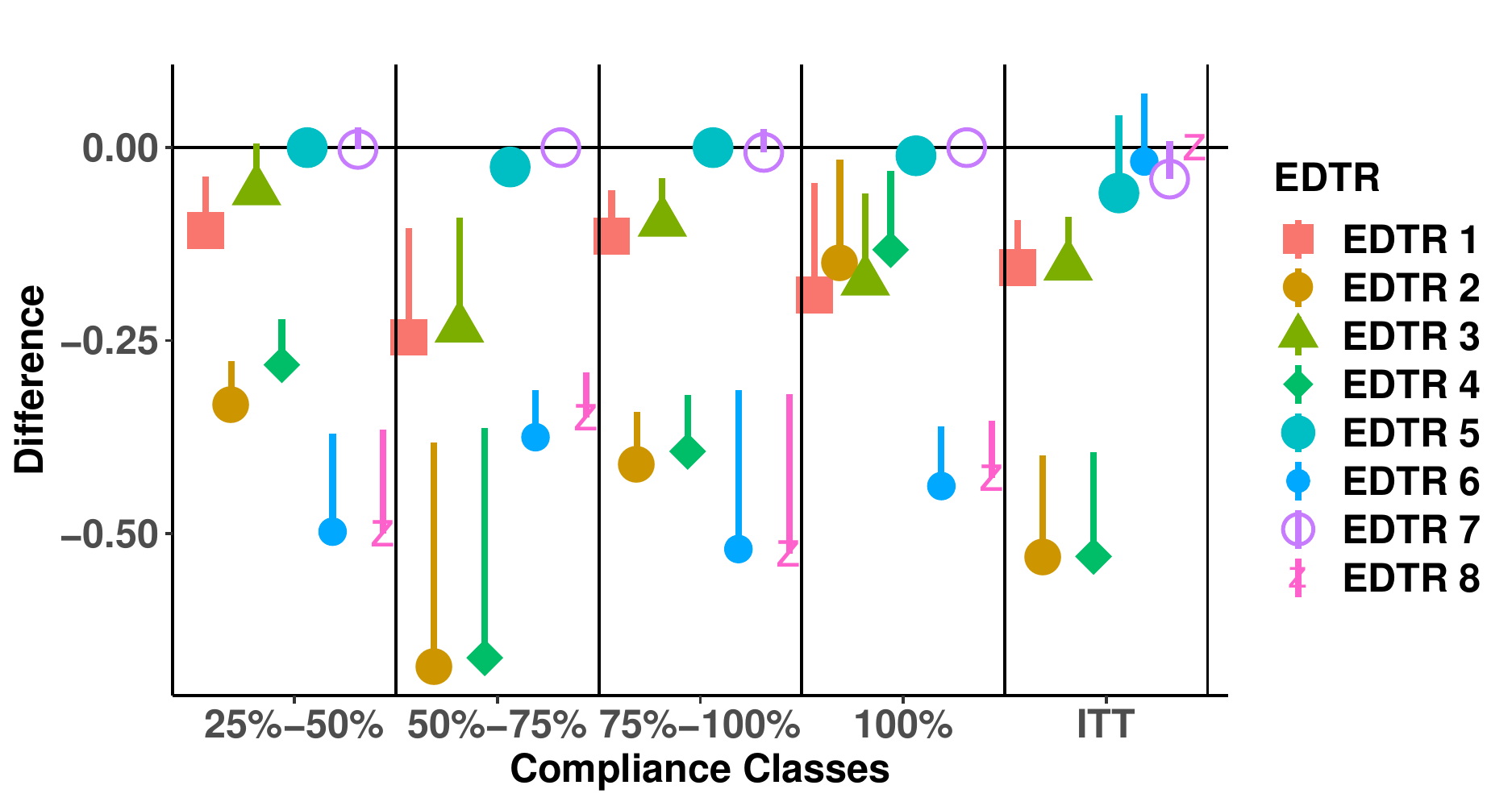}
\caption{EDTR = embedded DTR, ITT = intention-to-treat. Simultaneous upper one-sided credible intervals by compliance class for the difference between each EDTR and the optimal embedded DTR for main-effects models. Top graph: simulated ENGAGE SMART.Lower graph: general SMART simulation.
ITT are confidence intervals computed using the bootstrap.}
\label{fig:MCB-EDTR-Sim}
\end{figure}

\end{document}